\RequirePackage{fix-cm}
\documentclass[smallextended]{svjour3}       
\smartqed  
\usepackage{graphicx}

\usepackage{lipsum}

\usepackage{amsmath}
\usepackage{amsfonts}
\usepackage{amssymb}
\usepackage{mathdots}
\usepackage{mathabx}
\usepackage{mathdots}
\usepackage[utf8]{inputenc}
\usepackage{enumerate}
\usepackage{amsmath}
\usepackage{amssymb}
\usepackage{centernot}
\usepackage[mathscr]{euscript}
\usepackage{url}
\usepackage{hyphenat}
\usepackage{bm}
\usepackage{verbatim}
\usepackage{graphicx}
\usepackage{subcaption}
\usepackage{listings}
\usepackage{xcolor}
\usepackage{algorithm,algorithmicx}
\usepackage{algpseudocode}
\algnewcommand\algorithmicinput{\textbf{Input: }}
\algnewcommand\algorithmicoutput{\textbf{Output: }}
\usepackage{afterpage}
\usepackage{epstopdf}
\usepackage{cuted}
\usepackage{multicol}
\usepackage[title,toc,titletoc,header]{appendix}

\everymath{\displaystyle}

\newcommand\black[1]{\textcolor{black}{#1}}

\newcommand{\param}{\lambda}
\newcommand{\pspace}{\mathbf{\Lambda}}
\newcommand{\qspace}{\mathbf{\mathcal{D}}}
\newcommand{\qmap}{Q}

\newcommand{\pmeas}{\mu_{\pspace}}

\newcommand{\priordens}{\pi_{\pspace}^{\text{init}}}
\newcommand{\postdens}{\pi_{\pspace}^{\text{up}}}

\newcommand{\pfpriordens}{\pi_{\qspace}^{\qmap(\text{init})}}
\newcommand{\pfpostdens}{\pi_{\qspace}^{\qmap(\text{up})}}

\newcommand{\obsdens}{\pi_{\qspace}^{\text{obs}}}

\usepackage{lineno,hyperref}

\modulolinenumbers[5]

\usepackage{lineno}

\begin{document}

\title{Solving stochastic inverse problems for property-structure linkages using data-consistent inversion and machine learning
\thanks{This research was supported by the U.S. Department of Energy, Office of Science, Early Career Research Program.}
}

\titlerunning{Solving stochastic inverse problems for property-structure linkages}        

\author{Anh Tran \and
        Tim Wildey 
}


\institute{Anh Tran \thanks{Corresponding author: anhtran@sandia.gov}, Tim Wildey \at
              Optimization and Uncertainty Quantification Department \\
              Center for Computing Research \\
              Sandia National Laboratories \\
              \email{anhtran@sandia.gov}           
}

\date{Received: date / Accepted: date}

\maketitle

\begin{abstract}
Determining process-structure-property linkages is one of the key objectives in material science, and uncertainty quantification plays a critical role in understanding both process-structure and structure-property linkages.
In this work, we seek to learn a distribution of microstructure parameters that are consistent in the sense that the forward propagation of this distribution through a crystal plasticity finite element model (CPFEM) matches a target distribution on materials properties.
This stochastic inversion formulation infers a distribution of acceptable/consistent microstructures, as opposed to a deterministic solution, which expands the range of feasible designs in a probabilistic manner.
To solve this stochastic inverse problem, we employ a recently developed uncertainty quantification (UQ) framework based on push-forward probability measures, which combines techniques from measure theory and Bayes’ rule to define a unique and numerically stable solution.
This approach requires making an initial prediction using an initial guess for the distribution of model inputs and solving a stochastic forward problem.
To reduce the computational burden in solving both stochastic forward and stochastic inverse problems, we combine this approach with a machine learning (ML) Bayesian regression model based on Gaussian processes and demonstrate the proposed methodology on two representative case studies in structure-property linkages.
\end{abstract}

\section{Introduction}
\label{sec:Intro}


Discovering, designing, developing and manufacturing advanced materials through the process-structure-property relationship is one of the grand challenge problems in materials science~\cite{national2011materials}.
Such problems are typically challenging to rigorously solve due to the stochastic nature of microstructures, the high-dimensionality of feature spaces and/or observational data, the multiscale and time-dependent behavior of materials, and the scarcity of informative data.
However, the potential benefits for science and society are irrefutable.
To accelerate materials design and development, multiple integrated computation materials engineering (ICME) models have been proposed and developed over the last few decades to predict materials behaviors in practical settings.
Uncertainty quantification (UQ) plays a key role in establishing process-structure-property relationships due to the natural randomness of microstructure, measurement and model-form errors and parametric uncertainty in ICME models.
Many, but certainly not all, high-fidelity ICME models require substantial computational resources to make a single deterministic prediction.
Thus, the utilization of these high-fidelity models, especially in the context of optimization and uncertainty quantification, is often constrained by the available computational resources.
Machine learning (ML), including deep learning, is widely regarded as the fourth-paradigm in science, complementing experimental/empirical science, model-based theoretical science, and computational science~\cite{agrawal2016perspective}.
In the context of materials design and development, the utilization of ML to mimic high-fidelity ICME models while requiring significantly less computational resources
has tremendous potential to improve the overall efficiency of optimization and UQ.
To that end, we seek to solve a stochastic inverse problem in the context of inverse property-structure linkages by combining a forward ML model bridging structure-property linkages and a recently developed UQ framework based on data-consistent stochastic inversion.
More precisely, our goal is to infer a probability density function of microstructure features, such that the induced push-forward density of materials properties obtained from the ML surrogate model matches a given target density of materials properties.

Given the critical importance of optimization and UQ for a wide variety of problems in materials science, several frameworks have been developed over the years, see e.g., \cite{panchal2013key,mcdowell2007simulation,kalidindi2016vision}, to provide robust predictions under uncertainty.
A comprehensive review of UQ applications in ICME-based simulations can be found in Honarmandi and Arr{\'o}yave \cite{honarmandi2020uncertainty} and a comparison of different ML methods for predicting homogenized materials properties can be found in Fernandez-Zelaia et al~\cite{fernandez2019comparative}.
Generally speaking, homogenization problems are easier to solve because the number of quantities of interest (QoI) is usually relatively small, sometimes it is a single-output, as opposed to multi-output quantities that are often of interest in localization problems.
More specific to the work in this paper,
Paul et al.~\cite{paul2019microstructure} sought to optimize different materials properties to give sets of optimal and sub-optimal microstructures characterized by an orientation distribution function.
Johnson and Arr{\'o}yave~\cite{johnson2016inverse} proposed an inverse design framework for process-structure linkage and identified the heat treatment in Ni-rich NiTi shape memory alloys with a target size distribution of Ni$_4$Ti$_3$ precipitates.
Yuan et al.~\cite{yuan2018machine} used a crystal plasticity finite element method (CPFEM) to generate a training dataset and employed principal component analysis along with random forests to predict the stress-strain behavior.
Diehl et al.~\cite{diehl2017identifying} proposed an ICME workflow coupling DAMASK and DREAM.3D, which we also adopt in this work,
that was subsequently employed in Liu et al \cite{liu2020strategy,liu2020microstructure} and Diehl et al.~\cite{diehl2019quantifying} to quantify the influence of grain shape and crystallographic orientation of fine-structure dual-phase and high-strength low-alloy steel respectively.
A Gaussian process (GP) regression model and a Materials Knowledge System framework were combined in Tallman et al.~\cite{tallman2019gaussian,tallman2020uncertainty} to model a set of homogenized materials properties with respect to an orientation distribution function.
Liu et al. \cite{liu2015machine,liu2017context} developed a physics-based microstructure descriptors approach to parameterize microstructures as inputs and constructed the structure-properties map for a localization problem using regression trees and support vector machines.
Recently, more advanced deep learning techniques have been proposed to solve localization problems~\cite{yang2019establishing,pandey2020machine} and homogenization problems~\cite{yang2018deep,yang2019deep}, respectively.
Optimization and ML tools were also used in Wang and Adachi~\cite{wang2019property} for designing steel.
In a closely related work, Acar et al.~\cite{acar2017stochastic} proposed a linear programming approach to maximize a mean of materials properties under the assumption of Gaussian distribution for both inputs and outputs.
The problem we considered in this paper is somewhat similar to the work of Acar et al.~\cite{acar2017stochastic},
even though we do not assume a parametric representation, e g., a Gaussian distribution, for the target or inverse distributions and we do not pursue a deterministic optimization approach.
\black{Inductive design exploration method (IDEM) \cite{ellis2017application,mcdowell2009integrated,choi2008inductive} has been introduced as a materials design methodology to identify feasible and robust design for microstructure features, which has been applied to many problems in practice. 
The framework also formulates as a deterministic optimization problem, which finds a unique and optimal microstructure features, but currently does not allow incorporation of prior and posterior density function. Furthermore, the objectives of IDEM and the proposed framework also differ, in the sense that the previous one finds a unique microstructure features, while the later finds a consistent probability density function of microstructure features. 
}

In this paper, we seek to provide a probabilistic representation of the acceptable/consistent microstructure features.
This representation could be used to enhance the manufacturability of materials, but this is beyond the scope of this paper.
We utilize a recently developed formulation for solving stochastic inverse problems to generate samples from a probability density function of microstructure features that is model/data-consistent in the sense that the subsequent forward propagation of these samples matches a target probability density function of homogenized materials properties.
Most of the previously mentioned works considering materials designs have been cast into a deterministic optimization framework, which may able to search for the globally optimal microstructures that produce optimal homogenized materials properties.
While these previous deterministic approaches using optimization can incorporate aleatoric uncertainty in the objective function,
the deterministic formulation makes it very challenging to manufacture and produce the optimal microstructure in practice due to the fact that the homogenized materials properties do exhibit some variability, at least in the form of aleatory uncertainty due to spatial variation.
In our opinion, the inversion framework should be cast directly to the stochastic process-property relationship. 
While materials design considering uncertainty and microstructure sensitivity have been studied extensively, to the best knowledge of the authors, we are unaware of any prior work that solves a class of stochastic inverse problems in structure-property relationship using ML and UQ simultaneously and without assuming any parametric forms for the solutions.

The rest of this paper is organized as follows.  Section \ref{sec:MLcpfem} summarizes the ML workflow for forward structure-property predictions using DREAM.3D as a synthetic microstructure generation code, DAMASK as a CPFEM model and a GP regression model as an ML tool to bridge the structure-property relationship.
Section~\ref{sec:StochInv} provides background details about the stochastic inverse problems in UQ context and a straightforward  numerical implementation based on Monte Carlo sampling.
In Section~\ref{sec:casestudy-TWIP}, we present the first case study for twinning-induced plasticity (TWIP) steels, where the density of average grain size is inferred based on a ML surrogate for the Hall-Petch relationship to match target distribution of yield stress. 
Section~\ref{sec:casestudy-Al} presents the second case study for aluminum alloy, where the density of grain aspect ratio is inferred, using phenomenological constitutive model.
Section~\ref{sec:Discussion} provides some additional discussion of the results in the paper.
Finally, Section~\ref{sec:Conclusion} contains our concluding remarks.

\section{A forward UQ framework employing ML for the structure-property map using CPFEM and data mining}
\label{sec:MLcpfem}

In materials science, microstructures are often characterized by microstructure descriptors,
and due to the stochastic nature of microstructures, statistical microstructure descriptors are much more widely used than deterministic ones.
Typical examples of deterministic microstructure descriptor include volume fraction, total number of grains, and total number of surface areas, while statistical microstructure descriptors include, but are not limited to, two-point correlation function, chord-length distribution, grain size distribution, orientation distribution function, misorientation distributed function.
Interested readers are referred to the works of Groeber et al. \cite{groeber2008framework1,groeber2008framework2}, Bostanabad et al. \cite{bostanabad2018computational}, and Torquato \cite{torquato2002statistical} for comprehensive reviews of computationally characterizing microstructures. 

One approach for generating synthetic microstructures employs deterministic optimization to minimize differences of many microstructure descriptors.
Such approaches have been well studied in the literature, as reviewed by Bargmann et al~\cite{bargmann2018generation} and Liu et al~\cite{liu2013computational}, and even applied to experimental microstructures~\cite{tran2019data}.
Notably, even though statistical microstructure descriptors are mathematically represented as a probability density function, in practice, statistical microstructure descriptors are often parameterized by fitting over a parameterized family of probability density functions or discretized spatially into a finite number of bins, i.e., a histogram.
Examples families commonly used for parametric density fitting include Gaussian, log-normal, and Weibull distributions, where the probability density functions are completely controlled by a small number of parameters. 
In either cases, this effectively converts the representation and parameterization of deterministic or statistical descriptors into a vector of parameters, typically real-valued, which we generally denote as $\param$ in this paper.
The vector $\param \in \pspace \subset \mathbb{R}^k$ encodes $k$ continuous microstructure features, which can also be thought of as microstructure descriptors.
For example, this $\param$ vector can encode a few location and scale parameters for well-known distributions, such as \black{Weibull, Gaussian, or $t$-distribution.}

Using $\param$ to denote microstructure features, one can encode the deterministic and statistical microstructure descriptors which characterize the statistics of microstructures, and subsequently reconstruct statistically equivalent microstructures through solving an optimization problem as described above.
In this paper, we employ DREAM.3D~\cite{groeber2014dream} to generate ensembles of statistically equivalent microstructures.
As demonstrated in Section~\ref{sec:casestudy-TWIP}, significant variability may be observed in a materials response due to the inherent variability in the underlying microstructure and this effect is most prevalent in cases with larger grains relative to the size of the stochastic volume element (SVE).
To account for this variability, for each realization of $\param$ an ensemble of SVEs \black{of number} $N_\text{SVE}$ are generated.
In this paper, we use $m^{(i)}, 1 \leq i \leq N_\text{SVE}$ to denote the microstructures.

Given these microstructures, a computational model bridging a structure-property linkage, such as DAMASK~\cite{roters2012damask,roters2019damask}, is employed to evaluate $m$ homogenized materials responses for each microstructure $m^{(i)}$ in the ensemble.
We use $\hat{\qmap}^{(i)} = \left( \hat{\qmap}_1(m^{(i)}), \dots, \hat{\qmap}_m(m^{(i)}) \right)$, where $\hat{\qmap}^{(i)} \in \mathbb{R}^m$ to denote the vector of these homogenized materials responses.
In this paper, we are interested in both the stochastic homogenized response and the ensemble average homogenized materials properties which are computed by averaging over a finite number of SVEs in a Monte Carlo manner as
\begin{equation}
\qmap (\param) \approx N_{\text{SVE}}^{-1} \sum_{i=1}^{N_\text{SVE}} \hat{\qmap}^{(i)} (\param),
\end{equation}
where $N_\text{SVE}$ is the number of SVE considered and $\hat{\qmap}^{(i)}$ is the homogenized materials properties obtained in the $i^\text{th}$ microstructure realization.
In the limit of infinite ensemble members, this gives a deterministic structure-property map.
For completeness, in Section~\ref{sec:casestudy-TWIP} we consider and compare the solutions to the stochastic inverse problem using both the deterministic map and the stochastic map.

To summarize, in this paper the vector $\param \in \pspace$ denotes the inputs, which are the microstructure features, and the outputs/QoI are $\qmap(\param) \in \qspace = \qmap(\pspace)$, which the homogenized materials properties or the ensemble average of these homogenized materials properties.
In theory, one can perform any forward or inverse UQ study by exhaustively sampling the microstructure feature vector $\param$ across the microstructure space $\pspace$.
However, given the significant computational cost in evaluating this map, we seek to construct a structure-property dataset and use this to build a surrogate model to approximate $Q(\lambda)$.
This surrogate model can then be exhaustively sampled without running ICME models which greatly reduces the computational efforts.
While any ML tools can be used for this surrogate, in this paper, we limit the scope of our study to the well-known Gaussian process (GP) regression model, \black{as demonstrated by Tallman et al. \cite{tallman2019gaussian,tallman2020uncertainty}.}

\begin{figure}[!htbp]
\centering
\includegraphics[width=0.75\textwidth,keepaspectratio]{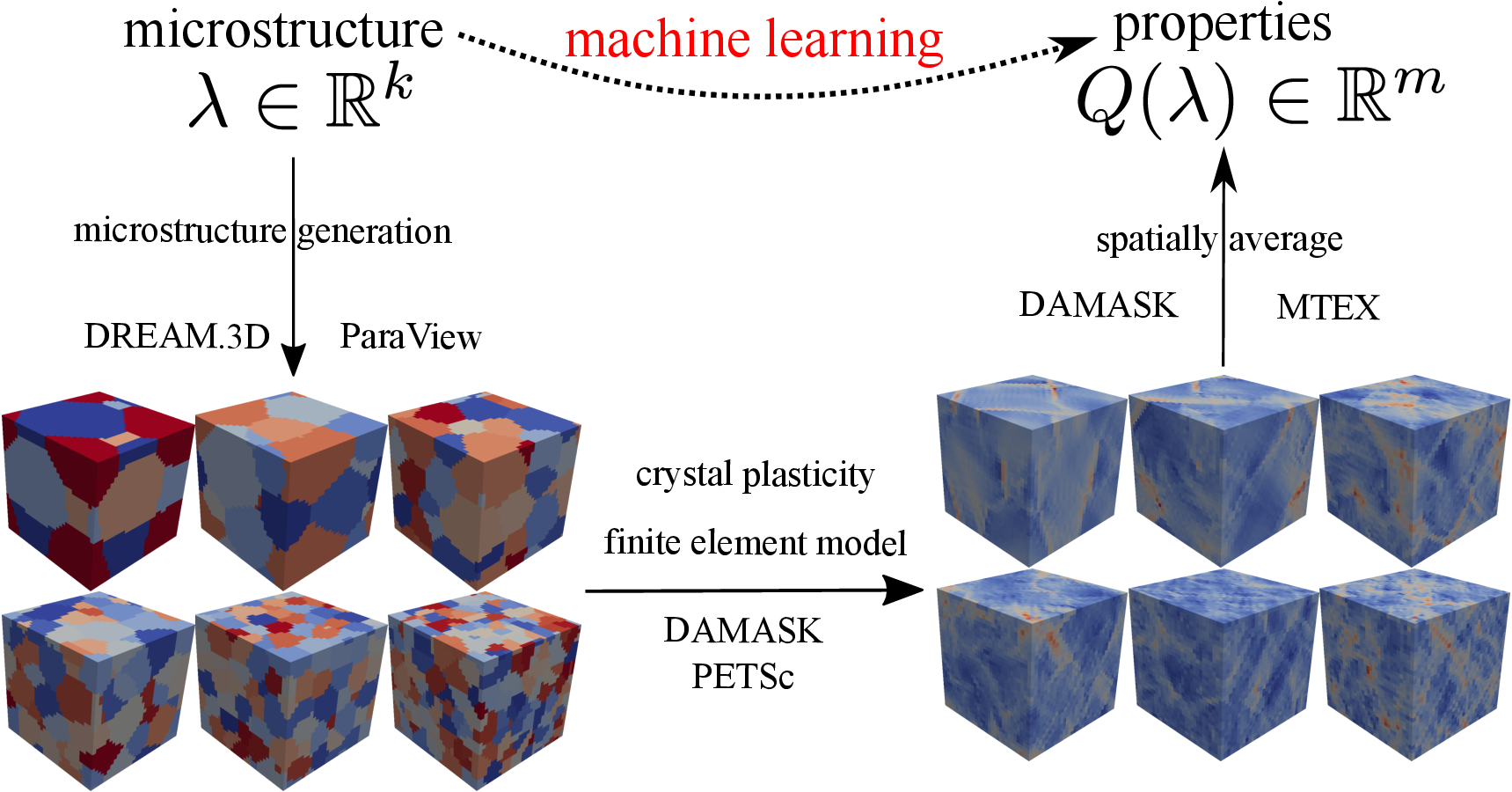}
\caption{Detailed illustration of the microstructure-homogenized materials properties map. The input is a $k$-dimensional microstructure feature vector $\lambda \in \mathbb{R}^k$.
For each realization of this vector, an ensemble of microstructure SVEs is generated using DREAM.3D,
and the homogenized material response is calculated using a CPFEM model, e.g., DAMASK.  Finally, machine learning is deployed to provide a surrogate approximation to the map $Q(\lambda)$ bridging the microstructure-homogenized materials properties relationship.}
\label{fig:ML-structure-property-scheme}
\end{figure}

In this paper, we adopt the workflow proposed by Diehl et al~\cite{diehl2017identifying} that couples a synthetic microstructure generation framework, here using DREAM.3D \cite{groeber2014dream}, and a forward ICME model, here a CPFEM model using DAMASK~\cite{roters2012damask,roters2019damask}, to bridge the structure-property relationship.
This automated workflow is used to explore the microstructure space, $\pspace$, while parallelization over and within the ensembles is utilized which exploits high-performance computing resources to generate the required datasets.
Figure~\ref{fig:ML-structure-property-scheme} summarizes the workflow mapping from microstructure $\param \in \pspace \subset\mathbb{R}^k$ -- a $k$-dimensional microstructure feature vector, to $\qmap(\param) \in \qspace\subset\mathbb{R}^m$ -- a $m$-dimensional materials properties vector, which are the QoIs.
In the rest of this section, Section~\ref{subsec:Dream3d} provides an overview of microstructure generation through DREAM.3D, while Section~\ref{subsec:cpfem} and Section~\ref{subsec:GP} summarize the basics of CPFEM and Gaussian process regression, respectively.

\subsection{Microstructure generation}
\label{subsec:Dream3d}

Synthetic microstructures are typically generated using a particle size distribution,
often defined as a log-normal distribution,
\begin{equation}
f_D(d; \mu_D, \sigma_D) = \frac{1}{ d \sigma_D \sqrt{2 \pi  }} e^{ - \frac{(\ln{d} - \mu_D)^2}{2\sigma_D^2} },
\end{equation}
where $\mu_D$ and $\sigma^2_D$ are the mean and the variance, respectively, of the normally distributed $\ln(D)$, i.e. $\ln(D) \sim \mathcal{N}(\mu_D, \sigma^2_D)$, and $D$ is the average grain diameter. 
The additional subscript in $\mu_D$ and $\sigma_D$ is introduced to avoid a conflict of notation between the mean and variance for the particle size distribution and the mean and variance for the GP model described in Section~\ref{subsec:GP}.
We employ the open-source DREAM.3D package~\cite{groeber2014dream} to generate synthetic microstructures from a given set of parameters $(\mu_D, \sigma_D)$ for the log-normal distribution.

\subsection{Crystal plasticity finite element model}
\label{subsec:cpfem}

In this section, we follow the notation and description from Roters et al.~\cite{roters2010overview,roters2011advanced,roters2011crystal,roters2019damask}, Diehl \cite{diehl2010spectral}, and Alharbi and Kalidindi \cite{alharbi2015crystal} for the constitutive relations in the CPFEM model.
For small deformations, the elasto-plastic decomposition can be computed additively, whereas for large deformations, a multiplicative decomposition is more appropriate,
\begin{equation}
\bm{F} = \bm{F}_\text{e} \cdot \bm{F}_\text{p},
\end{equation}
following by the elasto-plastic decomposition of the velocity gradient as
\begin{equation}
\bm{L} = \dot{\bm{F}} \cdot \bm{F}^{-1} = \dot{\bm{F}}_\text{e} \cdot \bm{F}_\text{e}^{-1} + \bm{F}_\text{e} \cdot \dot{\bm{F}}_\text{p} \cdot \bm{F}_\text{p} \cdot \bm{F}_\text{e}^{-1}
= \bm{L}_\text{e} + \bm{F}_\text{e} \cdot \bm{L}_\text{p} \cdot \bm{F}_\text{e}^{-1},
\end{equation}
where $\bm{L}_\text{p}$ and $\bm{L}_\text{e}$ are the plastic and elastic velocity gradient, respectively. 
The 2$^{\text{nd}}$ Piola-Kirchoff stress tensor $\bm{S}$, which is a symmetric second-order tensor defined in the intermediate configuration, is given by 
\begin{equation}
\bm{S} 
= \frac{\mathbb{C}}{2} : (\bm{F}_\text{e}^T \bm{F}_\text{e} - \bm{I}) 
= \mathbb{C} : \bm{E}_\text{e}
= J \bm{F}^{-1} \cdot \bm{\sigma} \cdot \bm{F}^{-T},
\end{equation}
where $\mathbb{C}$ is the elasticity fourth-order tensor, $\bm{F}_\text{e}$ is the elastic deformation gradient, $\bm{F}_\text{p}$ is the plastic deformation gradient \cite{roters2010overview}, $\bm{E}_\text{e} = \frac{1}{2}\left( \bm{F}^T_e \bm{F}_\text{e} - \bm{I} \right)$ is the elastic Green's Lagrangian strain, $\sigma$ is the Cauchy stress tensor (cf. \cite{roters2011crystal}, Section 3.3). 
Following Diehl et al~\cite{diehl2010spectral,eisenlohr2013spectral,shanthraj2015numerically}, we use the open-source DAMASK~\cite{roters2012damask,roters2019damask} package for the CPFEM model, and PETSc~\cite{abhyankar2018petsc,balay2019petsc} is used as the underlying spectral solver.

\subsection{Gaussian process regression}
\label{subsec:GP}

In this section, we adopt the notation from Shahriari et al~\cite{shahriari2016taking} and Tran et al~\cite{tran2019pbo,tran2020srmobo3gp,tran2020aphbo} for its clarity and consistency. 
In the context of this paper, a Gaussian process is a spatially distributed collection of random variables, each of which is normally distributed.
A $\mathcal{GP}(\mu_0,k)$ regressor is a nonparametric model
which is fully characterized by a prior mean function, $\mu_0(\param): \pspace \mapsto \mathbb{R}$, and a positive-definite kernel or a covariance function $k:\space \times \pspace \mapsto \mathbb{R}$.
In GP regression, it is assumed that the output, $f$, is jointly Gaussian, and the observations, $Q$, are normally distributed, leading to
\begin{equation}
\label{eq:prior}
f | \lambda \sim \mathcal{N}(m,\bm{K}),
\end{equation}
\begin{equation}
Q | f,\sigma^2 \sim \mathcal{N}(Q,\sigma^2 \bm{I}),
\end{equation}
where $m_i := \mu(\lambda_i)$, and $K_{i,j} := k(\lambda_i, \lambda_j)$. Equation \ref{eq:prior} describes the prior distribution induced by the GP. 

The Mat{\'e}rn family of kernels offer a broad range of options for stationary kernels, controlled by a smoothness parameter $\nu>0$ (cf. Section 4.2, \cite{rasmussen2006gaussian}), including the square-exponential ($\nu \to \infty$) and exponential kernels $\left( \nu = \frac{1}{2} \right)$ that are widely used in literature.
In this paper, we use the Mat{\'e}rn-3/2 kernel, where $\nu = \frac{3}{2}$, $k_{\text{Mat{\'e}rn}3} (\lambda, \lambda') = \theta_0^2 \exp{(-\sqrt{3}r)} (1+\sqrt{3} r)$, where $r$ is the distance between $\lambda$ and $\lambda'$.
Under these assumptions,
the prediction for an unknown arbitrary point is characterized by the posterior Gaussian distribution, which can be described by the posterior mean and posterior variance functions.
These are given by
\begin{equation}
\mu(\lambda) = \mu_0(\lambda) + \bm{k}(\lambda)^T (\bm{K} + \sigma^2 \bm{I})^{-1} (y - m),
\end{equation}
and
\begin{equation}
\sigma^2 (\lambda) = \bm{k}(\lambda) - \bm{k}(\lambda)^T (\bm{K}  + \sigma^2 \bm{I})^{-1} \bm{k}(\lambda),
\end{equation}
respectively, where $\bm{k}(\lambda)$ is a vector of covariance $\bm{k}(\lambda)_i = k(\lambda, \lambda_i)$, $\sigma^2 = \frac{1}{n} (y - \mu_0(\lambda))^T \bm{K}^{-1} (y - \mu_0(\lambda)) $ is the intrinsic variance.
To estimate the hyper-parameter $\theta = (\theta_i)_{i=1:d}$, we maximize the log marginal likelihood, which is computed as
\begin{equation}
\log p(y | \lambda_{1:n}, \theta) = - \frac{1}{2} (y - m)^T (\bm{K} + \sigma^2 \bm{I})^{-1} (y - m) - \frac{1}{2} \log{| \bm{K} + \sigma^2 \bm{I} |} - \frac{n}{2} \log{(2\pi)}.
\end{equation}
The main drawback of this surrogate representation is the scalability of computing the inverse of the covariance matrix, which is typically requires $\mathcal{O}(n^3)$ operations and memory.
For applications involving high-fidelity simulations, the scalability of GP regression is not a primary concern because $n$ is typically quite small.

\section{A stochastic inverse problem for structure-property relationship in materials design}
\label{sec:StochInv}

\subsection{Theory}

In this section, we provide a brief description of the data-consistent approach for solving a class of stochastic inverse problems introduced in~\cite{butler2018combining}.
For the ease of presentation, we follow the notation and terminology in~\cite{butler2018convergence} to emphasize the fact that this class of stochastic inverse problems is fundamentally different from the classical Bayesian formulation.
Interested readers are referred to~\cite{butler2018combining,butler2018convergence} for a longer discussion on these difference and for a precise mathematical formulation of the stochastic forward and inverse problems.

Recall that we use $\pspace \subset \mathbb{R}^k$ to denote a space of inputs and $\qspace \subset \mathbb{R}^m$ to denote the space of the QoI, which are connected via the parameter-to-QoI map $\qmap(\param):\pspace \to \qspace \subset \mathbb{R}^m$. 
\black{
The approach introduced in~\cite{butler2018combining} requires $\qmap$ to be a deterministic map (with stochastic inputs), but this was generalized in~\cite{butler2020stochastic} to incorporate stochastic maps.  We comment on the extension to stochastic maps at the end of this section.
}


Formally, the inverse problem is defined in terms of probability measures which is beyond the scope of this paper.
Here, we present the inverse problem in terms of probability density functions, since these are actually approximated and used in the algorithmic implementation, and refer the interested reader to~\cite{butler2018combining} for details.
Briefly stated, the stochastic inverse problem seeks a probability density on model input such that the forward propagation of this density through $\qmap$, often called a push-forward density, matches a given target probability density on the QoI.
Of course, the solution to this inverse problem is not necessarily unique, i.e., multiple probability densities may satisfy this requirement, and a solution may not exist if the model is not capable of producing the data. In~\cite{butler2018combining}, existence is guaranteed by a {\em predictibility assumption} and uniqueness and stability are obtained by a regularization technique similar to the use of a prior in Bayesian inference.
Following~\cite{butler2018combining}, we introduce an {\em initial} density, $\priordens(\param)$, on the model input parameters.  This initial measure/density are similar to a prior probability measure in classical Bayesian inference and can be chosen to incorporate prior knowledge or assumptions about the distribution on the input parameters.
We also define the {\em initial prediction} to be the push-forward of the initial density through the computational model.
We use $\pfpriordens$ to denote the corresponding density respectively on $\qspace$.
In Section~\ref{subsec:implementation}, we discuss how to approximate $\pfpriordens$ using standard UQ techniques, e.g., Monte Carlo sampling, rejection sampling and kernel density estimation.  More advanced techniques can also be used, but these standard approaches will be sufficient for this work.

Given the initial density, $\priordens$, and the corresponding push-forward density, $\pfpriordens$, we can define
an {\em updated} probability density, $\postdens$, that is a unique and numerically stable solution to the inverse problem and is given by
\begin{equation}\label{eq:updens}
\postdens(\param) = \priordens(\param) \frac{ \obsdens(\qmap(\param)) }{ \pfpriordens(\qmap(\param)) },
\quad \param \in \pspace.
\end{equation}
While the expression in~\eqref{eq:updens} resembles the posterior density in classical Bayesian inference, it is fundamentally different through the incorporation of the push-forward density.

\begin{figure}[!htbp]
\centering
\includegraphics[width=0.75\textwidth,keepaspectratio]{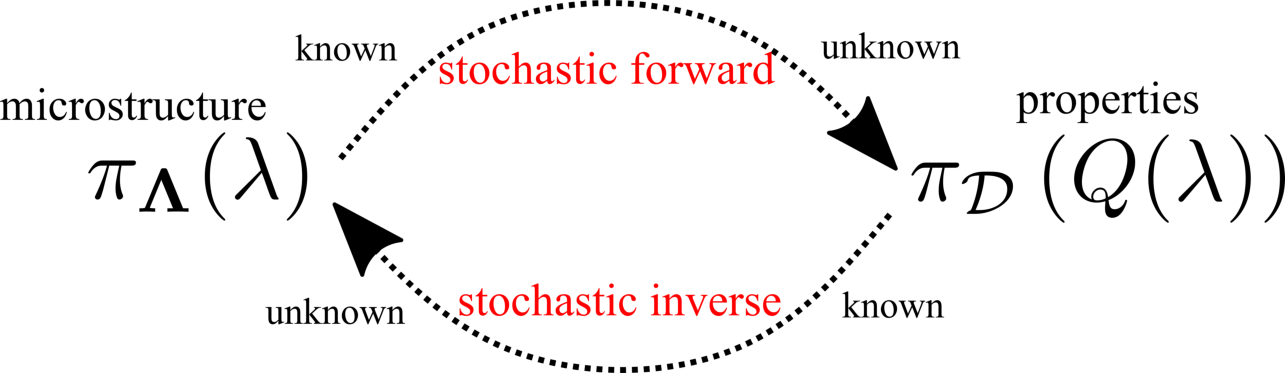}
\caption{Schematic illustrating the relationship between the stochastic forward and stochastic inverse problems.
In the context of structure-property relationships, the stochastic forward problem assumes that the distribution of inputs is known and seeks to solve for the distribution of outputs, while the inverse problem assumes a target distribution on the outputs is known and seeks a distribution on input that propagates forward to the target.
}
\label{fig:stochasticInverse.eps}
\end{figure}

\black{
Figure~\ref{fig:stochasticInverse.eps} illustrates the relationship between the stochastic forward and inverse problems, where the probability densities $\pi_{\pspace}(\param)$ and $\pi_{\qspace}(\qmap(\param))$ are associated with microstructure and properties, respectively.
The stochastic forward problem seeks to solve for the distribution on $\qmap(\param)$, given an assumed distribution on $\param$, while the stochastic inverse problem assumes a target distribution on $\qmap(\param)$ is given and seeks a consistent distribution on $\param$.}
In the context of the structure-property relationships, the stochastic inverse problem assumes that there is a desired/target distribution on the properties and seeks a distribution on the microstructure features such that the forward propagation of this distribution through the model matches the desired/target distribution.

As previously mentioned, this approach was extended in~\cite{butler2020stochastic} to incorporate stochastic maps.  Theoretically and algorithmically, the approach is very similar to the one described here and differs primarily in the interpretation of the updated density in $\pspace$ as a marginal of a consistent density in the joint space defined by $\pspace$ and the stochastic variability in the model.  We refer the interested reader to~\cite{butler2020stochastic} for details.


\subsection{Implementation and diagnostics}\label{subsec:implementation}

\begin{algorithm}[!htbp]
\caption{Acceptance-rejection sampling.}
\label{alg:rejectionSampling}
\algorithmicinput Probability densities $f(\param)$, $g(\param)$, $M>0$ such that $f(\param) \leq M g(\param)$ for all $\param \in \pspace$

\algorithmicoutput $N$ samples from $f(\param)$
\begin{algorithmic}[1]
\State $i \gets 0$
\While{$i < N$}
\State draw $\param^{(i)} \sim g(\param)$
\State draw $u \sim U(0,1)$
\If{$u < \frac{f(\param^{(i)})}{M \cdot g(\param^{(i)})}$}
\State accept $\param^{(i)}$; i++ \Comment{accept; increase $i$}
\Else
\State reject $\param^{(i)}$ \Comment{reject}
\EndIf
\EndWhile

\end{algorithmic}
\end{algorithm}

In practice, direct numerical construction of the updated density, $\postdens(\param)$, is impractical, so we often seek to generate a set of samples from this distribution.
A straightforward approach for generating these samples is rejection sampling.
For the sake of completeness, we briefly summarize the rejection sampling algorithm in Algorithm~\ref{alg:rejectionSampling}.
The main objective of Algorithm~\ref{alg:rejectionSampling} is to generate samples the target density $f(\param)$, using a proposal distribution $g(\param)$.
The acceptance-rejection algorithm continuously draws samples from $g(\param)$ and accept the samples under certain conditions, as described by Algorithm~\ref{alg:rejectionSampling}.
A key assumption in Algorithm~\ref{alg:rejectionSampling} is the existence of a constant $M>0$ such that $f(\param) \leq M g(\param)$ for all $\param \in \pspace$.
To generate samples from the updated density, $\postdens(\param)$ using $\priordens(\param)$ as the proposal density, we require the existence of a constant $M>0$ such that
\begin{equation}
\postdens(\param) = \priordens(\param) \frac{ \obsdens(\qmap(\param)) }{ \pfpriordens(\qmap(\param)) } \leq M \priordens(\param),
\end{equation}
which is precisely the predictability assumption from~\cite{butler2018combining}.

We can gain more intuition and computational diagnostics by considering the ratio
\begin{equation}\label{eq:ratio}
r(\param) = \frac{\obsdens(\qmap(\param))}{\pfpriordens(\qmap(\param))} = \frac{\postdens(\param)}{\priordens(\param)}
\end{equation}
which serves as re-weighting of the samples from the initial distribution.
If we assume that we utilize $P$ samples from initial distribution, $\left\{ \param^{(i)}\right\}_{i=1}^P$ to construct the push-forward of the initial density (the initial forward UQ prediction), then we can re-use this information
to estimate $M \approx \max_{1\leq i\leq P} r(\param^{(i)})$ since we have already evaluated $\qmap$ for these samples..
Moreover, we can actually use this ratio evaluated at these samples to give a Monte Carlo estimates of the integral of the updated density,
\begin{equation}\label{eq:meanup}
\int_\pspace \postdens(\param) d\pmeas = \int_{\pspace} \priordens(\param) r(\param) d\pmeas \approx \frac{1}{P} \sum_{i=1}^P r(\param^{(i)}),
\end{equation}
and the Kullback-Leibler divergence between the initial and updated densities,
\begin{equation}\label{eq:kldiv}
\begin{array}{lll}
\text{KL}(\postdens || \priordens) &=& \int_\pspace \postdens(\param) \log \left(\frac{\postdens(\param)}{\priordens(\param)}\right) d\pmeas \\
&=& \int_\pspace r(\param) \log (r(\param)) \priordens(\param) d\pmeas \\
&\approx& \frac{1}{P}\sum_{i=1}^P r(\param^{(i)}) \log(r(\param^{(i)})).\\
\end{array}
\end{equation}
These diagnostics are useful since integral of the updated density provides a numerical validation of the predictibility assumption and the Kullback-Leibler divergence provides the relative entropy, sometimes also called the information gained, between the initial and updated densities.

In the scope of this paper, $\qmap(\param)$ is the map from the microstructure space $\pspace$ to the homogenized materials properties space $\qspace$.
The push-forward and updated densities are well-approximated using the standard kernel density estimation (KDE) method
if a sufficient number of samples can be generated.
\black{
However, generating samples from the map using the DREAM.3D and DAMASK workflow (or any other high-fidelity workflow) is computationally expensive, so we employ the Gaussian process regression technique described in Section~\ref{subsec:GP} to approximate this map using a limited number of samples.
This is justified by the fact that in~\cite{butler2018convergence}, we proved that the errors in the push-forward and updated densities are bounded by the error in the approximate map.
}

To summarize, given a target distribution of homogenized materials properties $\obsdens$, we seek to learn a distribution of the microstructure features $\postdens(\param)$, such that its corresponding push-forward distribution of materials properties match the target homogenized materials properties.
Algorithm \ref{alg:rejectionSamplingStochInvPS} summarizes the numerical implementation solving stochastic inverse problems in the structure-property relationship. 

\begin{algorithm}[!htbp]
\caption{Acceptance-rejection sampling algorithm for solving stochastic inverse problems in structure-property linkage.}
\label{alg:rejectionSamplingStochInvPS}
\algorithmicinput 
\begin{itemize}
\item $\qmap(\param)$: the ML/GP approximating the structure-property map
\item $\obsdens(\qmap(\param))$: target materials joint density
\item $\priordens(\param)$: the initial density for $\param$ in microstructure space $\pspace$
\end{itemize}

\algorithmicoutput at least $N$ samples $\{\param^{(i)}\}_{i=1}^{N}$ from the updated density $\postdens(\param)$
\begin{algorithmic}[1]
\State $m \gets 0$
\While{$m<N$}
\For{$i = 1, \dots, P$}
\State draw $\param^{(i)} \sim \priordens(\param)$
\EndFor
\State construct KDE approx. of $\pfpriordens(\qmap(\param))$ \Comment{forward UQ: $\qmap$ is the surrogate ML model}
\For{$i = 1, \dots, P$}
\State $r^{(i)} = r(\qmap(\param^{(i)}))$ \Comment{$r(\qmap(\param^{(i)}))$ defined in~\eqref{eq:ratio}}
\EndFor
\State $M \gets \max_{i} r^{(i)}$
\For{$i=1,\dots,P$}
\State draw $u \sim U(0,1)$
\If{$u < r^{(i)}/M$}
\State accept $\param^{(i)}$; $m++$ \Comment{accept; increase $m$}
\Else
\State reject $\param^{(i)}$ \Comment{reject}
\EndIf
\EndFor
\EndWhile
\end{algorithmic}
\end{algorithm}

\section{Case study 1: Equiaxed grains for TWIP steels under uniaxial tension}\
\label{sec:casestudy-TWIP}


TWIP steels have attracted significant attention lately due to their outstanding mechanical properties, including high strength and ductility.
In this case study, we employ a CPFEM model to computationally probe the Hall-Petch relationship and apply the proposed framework to find the probability density of average grain size that induces a push-forward probability density that matches a target probability density of yield stress.

\subsection{Constitutive models and materials parameters}\label{subsec_case1_model}

In this section, we consider Fe-22Mn-0.6C TWIP steel and adopt the dislocation-density-based constitutive model with material parameters as described in Steinmetz et al.~\cite{steinmetz2013revealing} and summarized in Section 6.2.3 and Tables 8 and 9 in \cite{roters2019damask}, respectively. The constitutive model was validated experimentally by Wong et al~\cite{wong2016crystal}, and for the sake of completeness, we briefly summarize the constitutive model here.
The TWIP/TRIP steel constitutive model is parameterized in terms of dislocation density, $\varrho$, the dipole dislocation density, $\varrho_{\text{di}}$, the twin volume fraction, $f_{\text{tw}}$, and the $\varepsilon$-martensite volume fraction, $f_{\text{tr}}$.
A model for the plastic velocity gradient with contribution of mechanical twinning and phase transformation was developed in Kalidindi~\cite{kalidindi1998incorporation} and is given by
\begin{equation}
\mathbf{L}_\text{p} = (1 - f^{\text{tot}}_{\text{tw}} -f^{\text{tot}}_{\text{tr}}) \sum_{\alpha=1}^{N_{\text{s}}} \dot{\gamma}^{\alpha} \mathbf{s}_{\text{s}}^{\alpha} \otimes \mathbf{n}_{\text{s}}^{\alpha} 
+ \sum_{\beta=1}^{N_{\text{tw}}} \dot{\gamma} \mathbf{s}_{\text{tw}}^{\beta} \otimes \mathbf{n}_{\text{tw}}^{\beta} 
+ \sum_{\chi=1}^{N_{\text{tr}}} \dot{\gamma}^{\chi} \mathbf{s}_{\text{tr}}^{\chi} \otimes \mathbf{n}_{\text{tr}}^{\chi},
\end{equation}
where $\chi = 1, \dots,N_{\text{tr}}$ is the $\varepsilon$-martensite with volume fraction $f_{\text{tr}}$ on $N_{\text{tr}}$ transformation systems, $\mathbf{s}_{\text{s}}^{\alpha}$ and $\mathbf{n}_{\text{s}}^{\alpha}$ are unit vectors along the shear direction and shear plane normal of $N_{\text{s}}$ slip systems $\alpha$, $\mathbf{s}_{\text{tw}}^{\beta}$ and $\mathbf{n}_{\text{tw}}^{\beta}$ are those of $N_{\text{tw}}$ twinning systems $\beta$, and $\mathbf{s}_{\text{tw}}^{\chi}$ and $\mathbf{n}_{\text{tr}}^{\chi}$ are those of $N_{\text{tr}}$ transformation systems $\chi$.
The Orowan equation models the shear rate on the slip system $\alpha$ as 
\begin{equation}
\dot{\gamma}^{\alpha} = \rho_e b_{\text{s}} \nu_0 \exp \left[ - \frac{Q}{k_B T} \left\{ 1 - \left( \frac{\tau_{\text{eff}}^{\alpha}}{\tau_{\text{sol}}} \right)^p \right\}^q \right],
\end{equation}
where $b_{\text{s}}$ is the length of the slip Burgers vector, $\nu_0$ is a reference velocity, $Q_{\text{s}}$ is the activation energy for slip, $k_B$ is the Boltzmann constant, $T$ is the temperature, $\tau_{\text{eff}}$ is the effective resolved shear stress, $\tau_{\text{sol}}$ is the solid solution strength, $0 < \rho_\text{s} \leq 1$ and $1 \leq q_\text{s} \leq 2 $ are fitting parameters controlling the glide resistance profile. 
Blum and Eisenlohr~\cite{blum2009dislocation} models the evolution of dislocation densities, particularly the generation of unipolar dislocation density and formation of dislocation dipoles, respectively, as
\begin{equation}
\dot{\varrho} = \frac{|\dot{\gamma}|}{b_\text{s}} \Lambda_\text{s} - \frac{2\hat{d}{b_\text{s}}} \varrho |\dot{\gamma}|, \quad
\dot{\varrho}_\text{di} = \frac{2(\hat{d} - \widecheck{d})}{b_\text{s}} \varrho |\dot{\gamma}| - \frac{2 \widecheck{d}}{b_\text{s}} \varrho_\text{di} |\dot{\gamma}| - \varrho_\text{di} \frac{4\nu_\text{cl}}{\hat{d} - \widecheck{d}},
\end{equation}
where the dislocation climb velocity is 
$\nu_\text{cl} = \frac{G D_0 V_\text{cl}}{\pi (1 - \nu) k_\text{B} T} \frac{1}{\hat{d} + \widecheck{d}}\exp \left( - \frac{Q_\text{cl}}{k_\text{B} T} \right)$,
$D_0$ is the pre-factor of self-diffusion coefficient, $V_\text{cl}$ is the activation volume for climb, $Q_\text{cl}$ is the activation energy for climb, $\hat{d} = \frac{3G b_\text{s}}{16\pi |\tau|}$ is the glide distance below which two dislocations form a stable dipole, $\widecheck{d} = D_a b_\text{s}$ is the distance below which two dislocations annihilate.
Strain hardening is described in terms of a dislocation mean free path, where the mean free path is denoted by $\Lambda$. 
The mean free path for slip is modeled as
\begin{equation}
\frac{1}{\Lambda_\text{s}} = \frac{1}{D} + \frac{1}{\lambda_\text{s}} + \frac{1}{\lambda_\text{tw}} + \frac{1}{\lambda_\text{tr}}
\end{equation}
where
\begin{equation}
\frac{1}{\lambda_\text{s}^{\alpha}} = \frac{1}{i_\text{s}} \left( \sum_{\alpha'=1}^{N_\text{s}} p^{\alpha \alpha'} (\varrho^{\alpha'} + \varrho_\text{di}^{\alpha'}) \right)^{1/2}, \quad
\frac{1}{\lambda_{\text{tw}}^{\alpha}} = \sum_{\beta = 1}^{N_{\text{tw}}} h^{\alpha \beta} \frac{f^\beta_{\text{tw}}}{t_\text{tw} (1 - f^\text{tot}_\text{tw})}, \quad
\frac{1}{\lambda_\text{tr}^{\alpha}} = \sum_{\chi=1}^{N_\text{tr}} h^{\alpha \chi} \frac{f^{\chi}_\text{tr}}{t_\text{tr}(1 - f^\text{tot}_\text{tr})},
\end{equation}
where $D$ is the average grain size, $i_\text{s}$ is a fitting parameter, $t_\text{tw}$ is the average twin thickness, and $t_\text{tr}$ is the average $\varepsilon$-martensite thickness. 
The mean free path for twinning and for transformation are computed, respectively, as
\begin{equation}
\frac{1}{\Lambda^\beta_\text{tw}} = \frac{1}{i_\text{tw}} \left( \frac{1}{D} + \sum_{\beta'=1}^{N_\text{tw}} h^{\beta \beta'} f_{\text{tw}}^{\beta'} \frac{1}{t_\text{tw} ( 1 - f^\text{tot}_\text{tw})} \right), \quad
\frac{1}{\Lambda^\chi_\text{tr}} = \frac{1}{i_\text{tr}} \left( \frac{1}{D} + \sum_{\chi'=1}^{N_\text{tr}} h^{\chi \chi'} f_{\text{tr}}^{\chi'} \frac{1}{t_\text{tr} ( 1 - f^\text{tot}_\text{tr})} \right),
\end{equation}
$i_\text{w}$ and $i_\text{tr}$ are fitting parameters. 
The nucleation rates for twins and $\varepsilon$-martensite are $\dot{N} = \dot{N}_0 P_\text{ncs} P$, where the probability $P$ that a nucleus bows out to form a twin or $\varepsilon$-martensite is 
\begin{equation}
p_\text{tw} = \exp \left[ - \left( \frac{\hat{\tau}_\text{tw}}{\tau} \right)^{p_\text{tw}} \right], \quad
p_\text{tr} = \exp \left[ - \left( \frac{\hat{\tau}_\text{tr}}{\tau} \right)^{p_\text{tr}} \right],
\end{equation}
$p_\text{tw}$ and $p_\text{tr}$ are fitting parameters. For more details, readers are referred to Roters et al. \cite{roters2019damask} (cf. Section 6.2.3) and Wong et al \cite{wong2016crystal}. 
Table \ref{tab:TWIPparams} lists the parameters used in this work. 

\begin{table}[!htbp]
\centering
\caption{TWIP constitutive model parameters (cf. Tables 8 and 9 in Roters et al. \cite{roters2019damask}).}
\label{tab:TWIPparams}
\begin{tabular}{|l|l|l|l|l|l|} \hline
\textbf{Property}              & \textbf{Value}      & \textbf{Unit} & \textbf{Property}                          & \textbf{Value}      & \textbf{Unit}  \\ \hline
$p_\text{tw}$                  & 5.0                 &               & $q_\text{s}$                               & 1.0                 &                \\
$p_\text{tr}$                  & 8.0                 &               & $D$                                        & vary                & $\mu m$        \\
$p_\text{s}$                   & 1.15                &               & $V_\text{cs}$                              & 1                   & $b_\text{s}^3$ \\
$\nu_0$                        & $10^{-4}$           & $ms^{-1}$     & $V_\text{cl}$                              & 1.5                 & $b_\text{s}^3$ \\
$D_\text{a}$                   & 2.0                 &               & $D_0$                                      & $4\cdot 10^{-5}$    & $m^2 s^{-1}$   \\
$i_\text{s}$                   & 30.0                &               & $b_\text{s}$                               & 256                 & pm             \\
$i_\text{tw}$                  & 10.0                &               & $b_\text{tw}$                              & 120                 & pm             \\
$i_\text{tr}$                  & 3.0                 &               & $b_\text{tr}$                              & 147                 & pm             \\
$t_\text{tw}$                  & 0.05                & $\mu m$       & $L_\text{tw}$                              & 0.192               & $\mu m$        \\
$t_\text{tr}$                  & 0.1                 & $\mu m$       & $L_\text{tr}$                              & 0.128               & $\mu m$        \\
$Q_\text{s}$                   & $3.5\cdot 10^{-19}$ & J             & $Q_\text{cl}$                              & $3.0\cdot 10^{-19}$ & J              \\ 
$\tau_\text{sol}$              & 0.15                & GPa           &                                            &                     & GPa            \\ 
$C_{11}$ ($\gamma$-austenite)  & 175.0               & GPa           & $C_{11}$ ($\varepsilon$-martensite)        & 242.3               & GPa            \\ 
$C_{12}$ ($\gamma$-austenite)  & 115.0               & GPa           & $C_{12}$ ($\varepsilon$-martensite)        & 117.7               & GPa            \\ 
$C_{13}$ ($\gamma$-austenite)  &                     & GPa           & $C_{13}$ ($\varepsilon$-martensite)        &  45.0               & GPa            \\ 
$C_{33}$ ($\gamma$-austenite)  &                     & GPa           & $C_{33}$ ($\varepsilon$-martensite)        & 315.0               & GPa            \\ 
$C_{44}$ ($\gamma$-austenite)  & 135.0               & GPa           & $C_{44}$ ($\varepsilon$-martensite)        &  40.5               & GPa            \\ \hline
\end{tabular}
\end{table}

\subsection{Forward UQ for materials behaviors}

\begin{figure}[!htbp]
\centering
\includegraphics[width=0.75\textwidth,keepaspectratio]{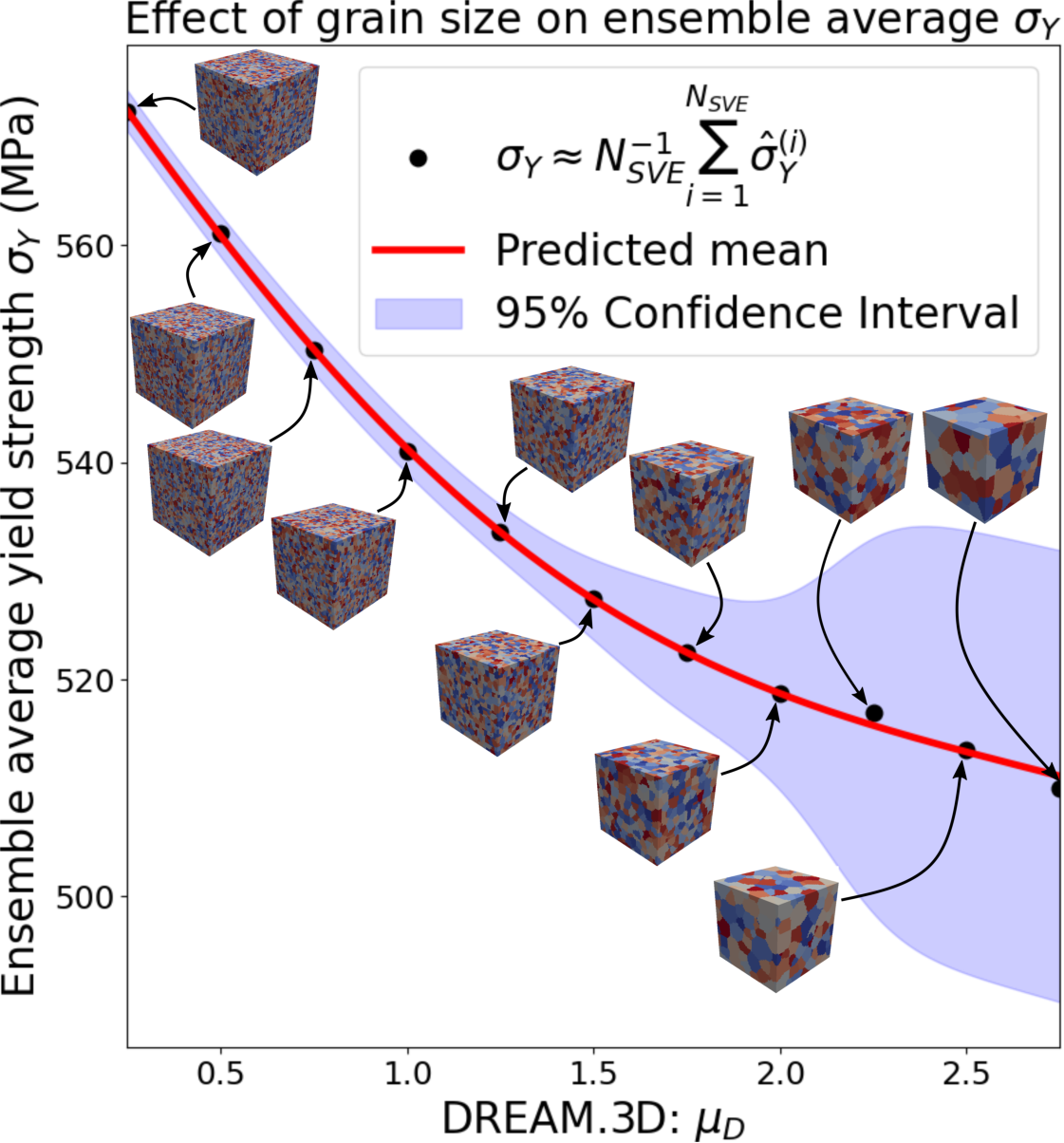}
\caption{The ensemble average homogenized yield stress $\sigma_Y$ versus the logarithm of average grain size $\mu_D$.
As the logarithm of average grain size grows larger, the ensemble-averaged yield stress decreases, but the variation increases due to fewer grains present in the SVE.
Here, we show the data points along with the heteroscedastic GP model, where the trend are captured in both the posterior mean and posterior variance. Highlighted region is the $95\%$
confidence intervals from the posterior variance of the GP model.
}
\label{fig:grainsizeDamaskDream3d}
\end{figure}

\begin{figure}[!htbp]
\begin{subfigure}[b]{0.475\textwidth}
\includegraphics[height=190px,keepaspectratio]{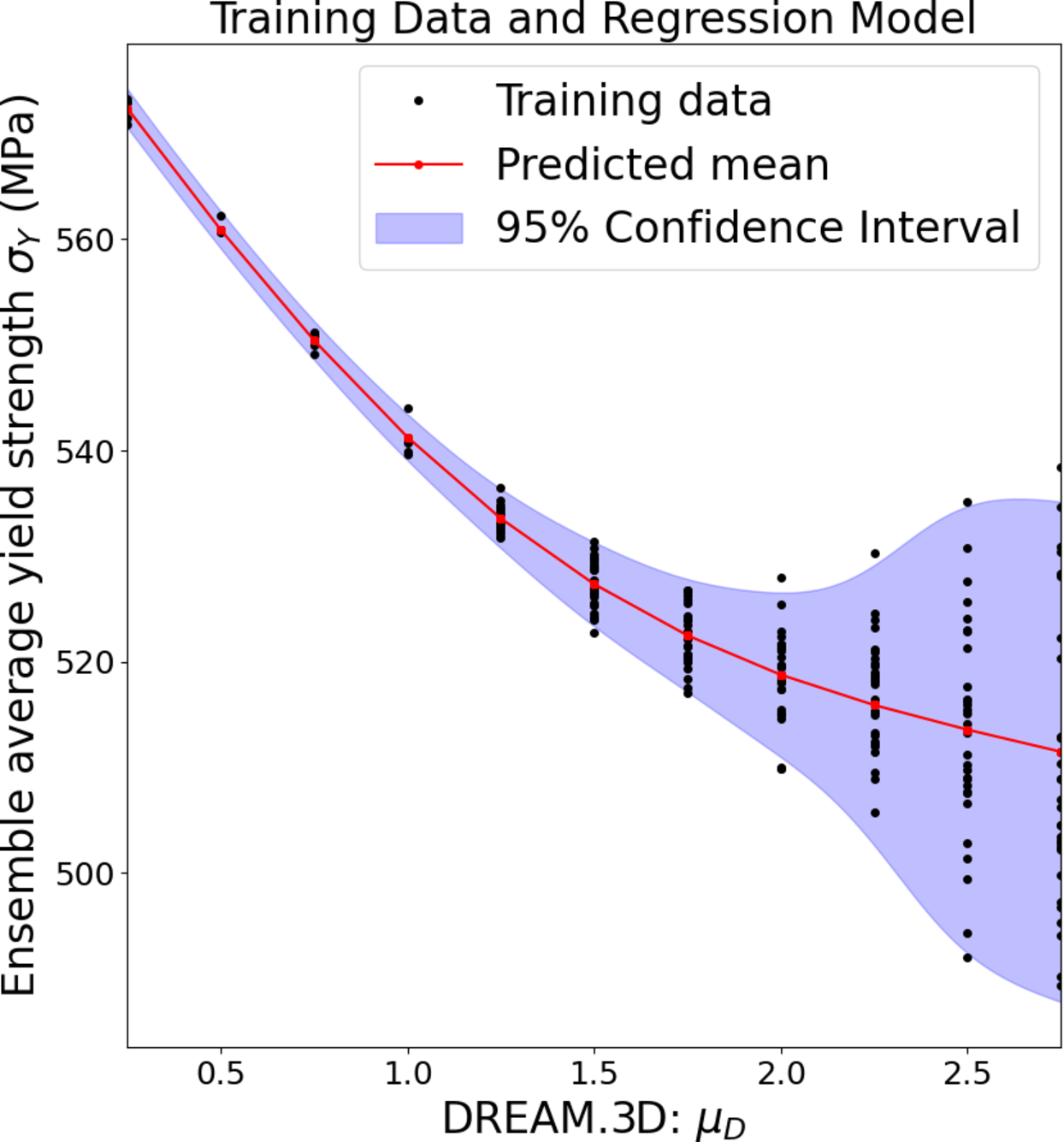}
\end{subfigure}
\hfill
\begin{subfigure}[b]{0.475\textwidth}
\includegraphics[height=190px,keepaspectratio]{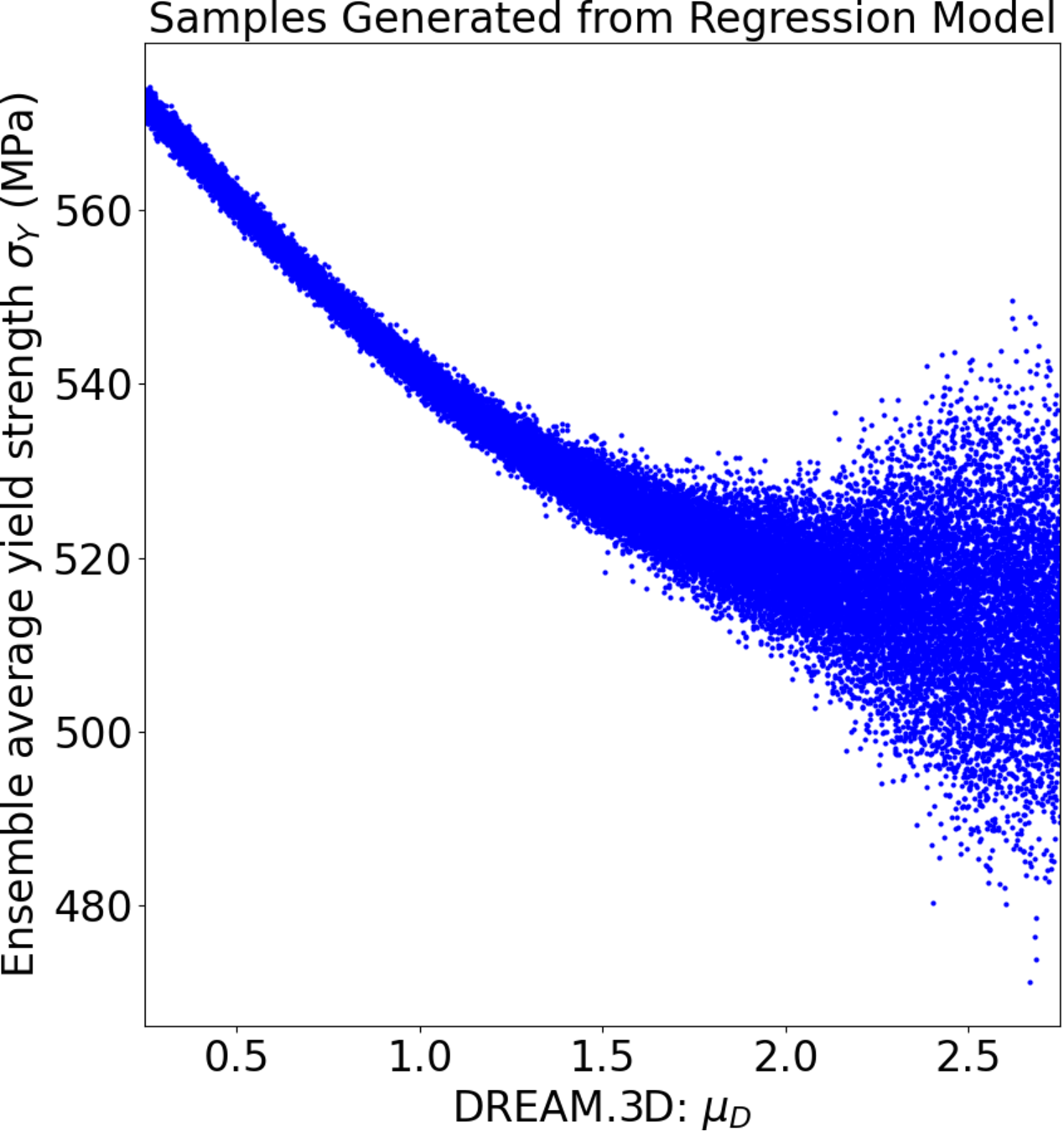}
\end{subfigure}
\caption{On the left, the ensemble average homogenized yield stress, $\sigma_Y$, as a function of the logarithm of the average grain size, $\mu_D$, along with the heteroscedastic GP regression model.
The highlighted region is the $95\%$ confidence intervals from the posterior varaince of the GP model.  On the right, the samples generated from the initial density and the GP regression model.
}
\label{fig:smap_training_predictions}
\end{figure}

In this case study, we vary the grain size parameter, $\mu_D$, in DREAM.3D and adjust the average grain size $D$ in DAMASK accordingly, as $D = e^{\mu_D}$.
The QoI is the offset yield stress $\hat{\sigma}_Y$ at $\varepsilon = 0.002$ under uniaxial tension with $\dot{\varepsilon}_{11} = 0.001 \text{s}^{-1}$.
An ensemble of 25 SVEs, each representing a $64\mu m \times 64\mu m \times 64 \mu m$ physical domain, is generated through DREAM.3D, where the CPFEM simulation is performed on a $64\times 64 \times 64$ grid.
The ensemble average homogenized yield stress is computed in a Monte Carlo manner, $\sigma_Y \approx N_{\text{SVE}}^{-1} \sum_{i=1}^{N_\text{SVE} = 25} \hat{\sigma}^{(i)}_Y $.  
Figure \ref{fig:grainsizeDamaskDream3d} shows the variation of $\sigma_Y$ as $\mu_D$ increases along with an example of SVE in the microstructure ensemble at each $\lambda = \mu_D$ considered. 
Figure~\ref{fig:smap_training_predictions} shows the homogenized response for each member of the ensemble, the mean of the regression model, the confidence interval, $\mu \pm 2\sigma$, and the samples generated from the initial density and the regression model.

\begin{figure}[!htbp]
\begin{subfigure}[b]{0.475\textwidth}
\includegraphics[height=190px,keepaspectratio]{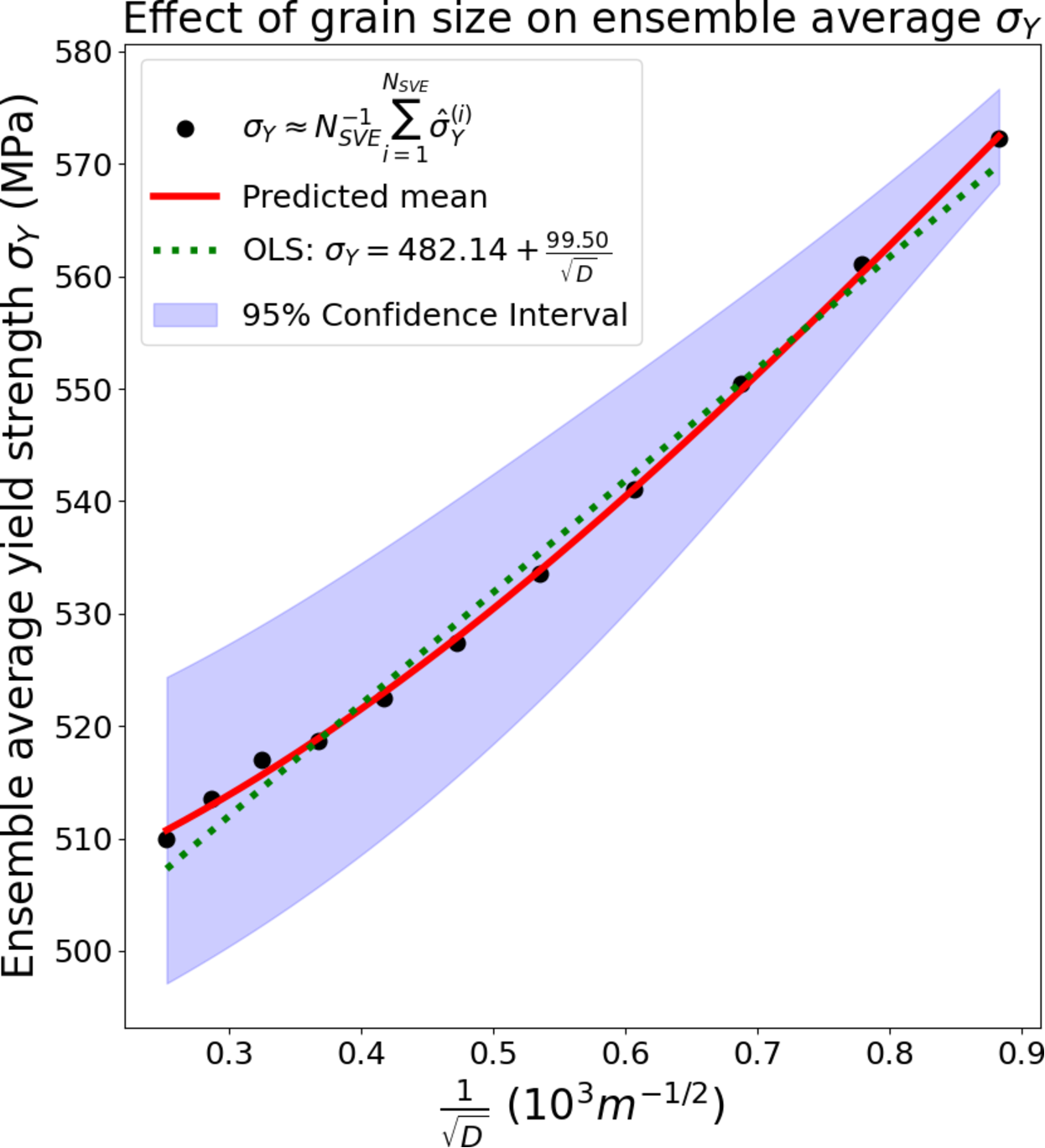}
\end{subfigure}
\hfill
\begin{subfigure}[b]{0.475\textwidth}
\centering
\includegraphics[height=190px,keepaspectratio]{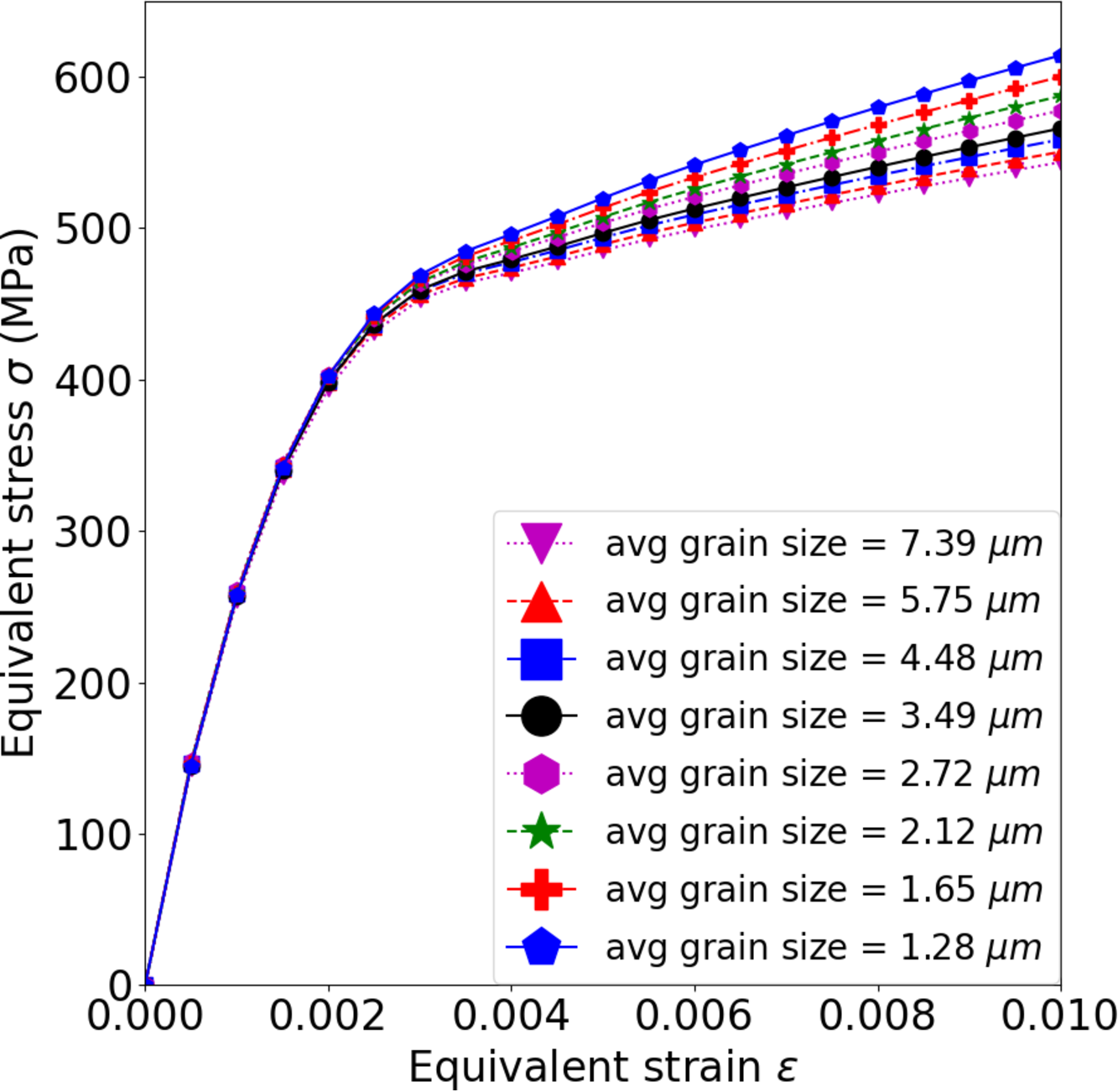}
\end{subfigure}
\caption{On the left, the inferred ML Hall-Petch relationship with 95\% confidence intervals for a TWIP steel, and a comparison with a least squares regression model.
On the right, the representative equivalent stress-strain relationship with respect to various average grain sizes. CPFEM simulations are performed with $64\mu m \times 64\mu m \times 64\mu m$ SVE at the loading condition of $\dot{\varepsilon}_{11} = 0.001 \text{s}^{-1}$.}
\label{fig:grainsizeHallPetch_SS_TWIP}
\end{figure}

Figure~\ref{fig:grainsizeHallPetch_SS_TWIP} (left) shows the inferred Hall-Petch relationship between $\sigma_Y$ and the inverse square root of the average grain size $D$ by transforming variables from Figure~\ref{fig:grainsizeDamaskDream3d}, along with the 95\% confidence intervals.
Figure~\ref{fig:grainsizeHallPetch_SS_TWIP} (right) shows a collection of representative stress-strain relationship with respect to different average grain sizes.
The parameter $\mu_D$ is sampled from 0.25 to 2.75, with a spacing of 0.25. 
Using linear regression, we obtain a fitted Hall-Petch relationship as $\sigma_Y = \sigma_0 + \frac{k}{\sqrt{D}}$, where \black{$\sigma_0 = 482.14$ MPa, and $k = 0.0995 \text{MPa} \ \text{m}^{1/2}$}. 
It is noted that increasing $\mu_D$ while keeping the same size of SVE leads to a smaller number of grains as $\mu_D$ increases. 
If the number of grains are denoted as $N_\text{grain}$, then $N_\text{grain}$ scales as $\mathcal{O}\left( e^{-3\mu_D} \right)$, which naturally leads to the observation that the noise is considerably large when $\mu_D$ reaches certain threshold.
We found that with the mesh of $64\times 64\times 64$, the error is acceptable with the upper bound of $\mu_D$ of 2.75. 

\subsection{Inverse UQ for microstructure features}

\begin{figure}[!htbp]
\begin{subfigure}[b]{0.475\textwidth}
\centering
\includegraphics[height=190px,keepaspectratio]{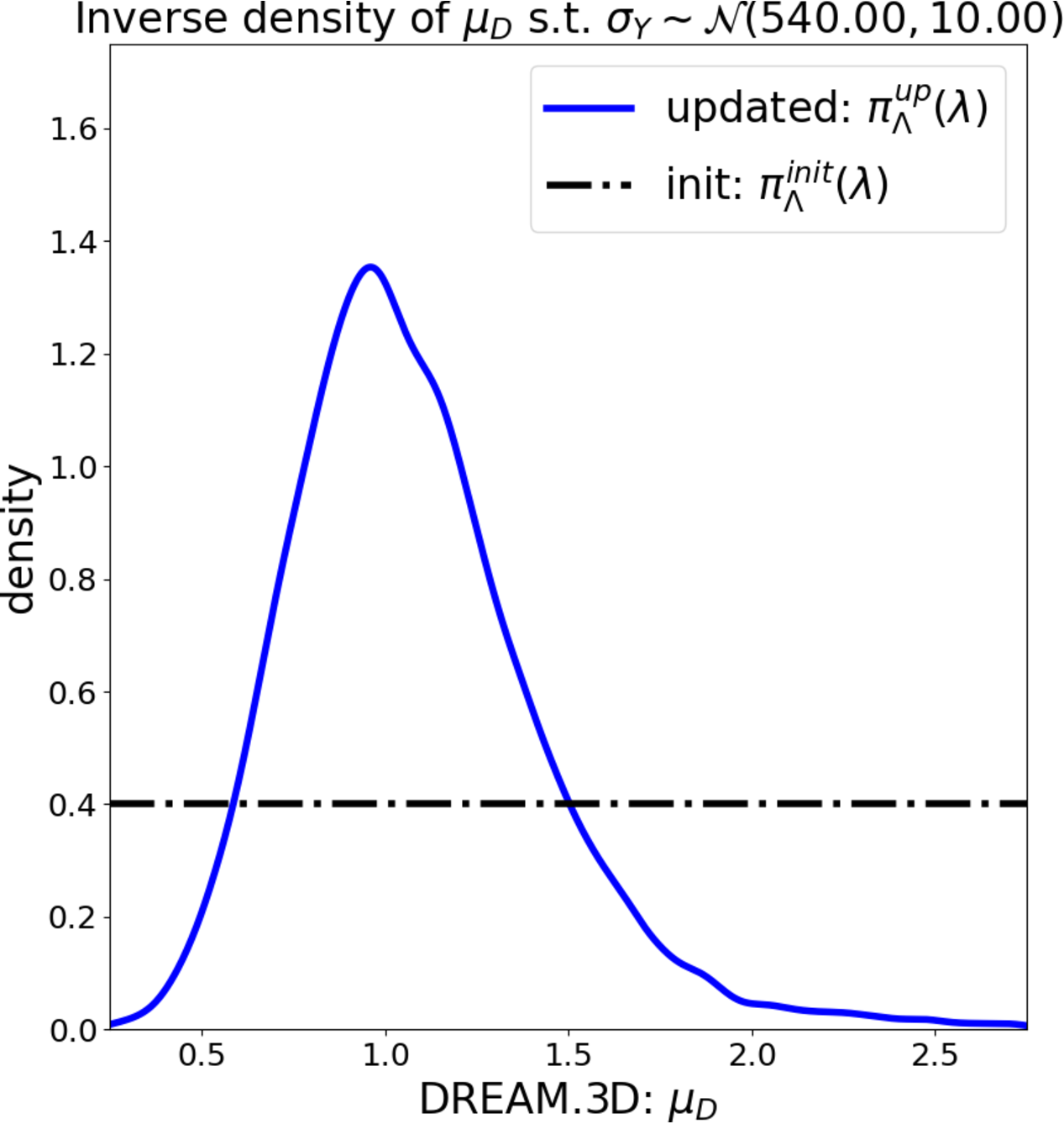}
\end{subfigure}
\hfill
\begin{subfigure}[b]{0.475\textwidth}
\centering
\includegraphics[height=190px,keepaspectratio]{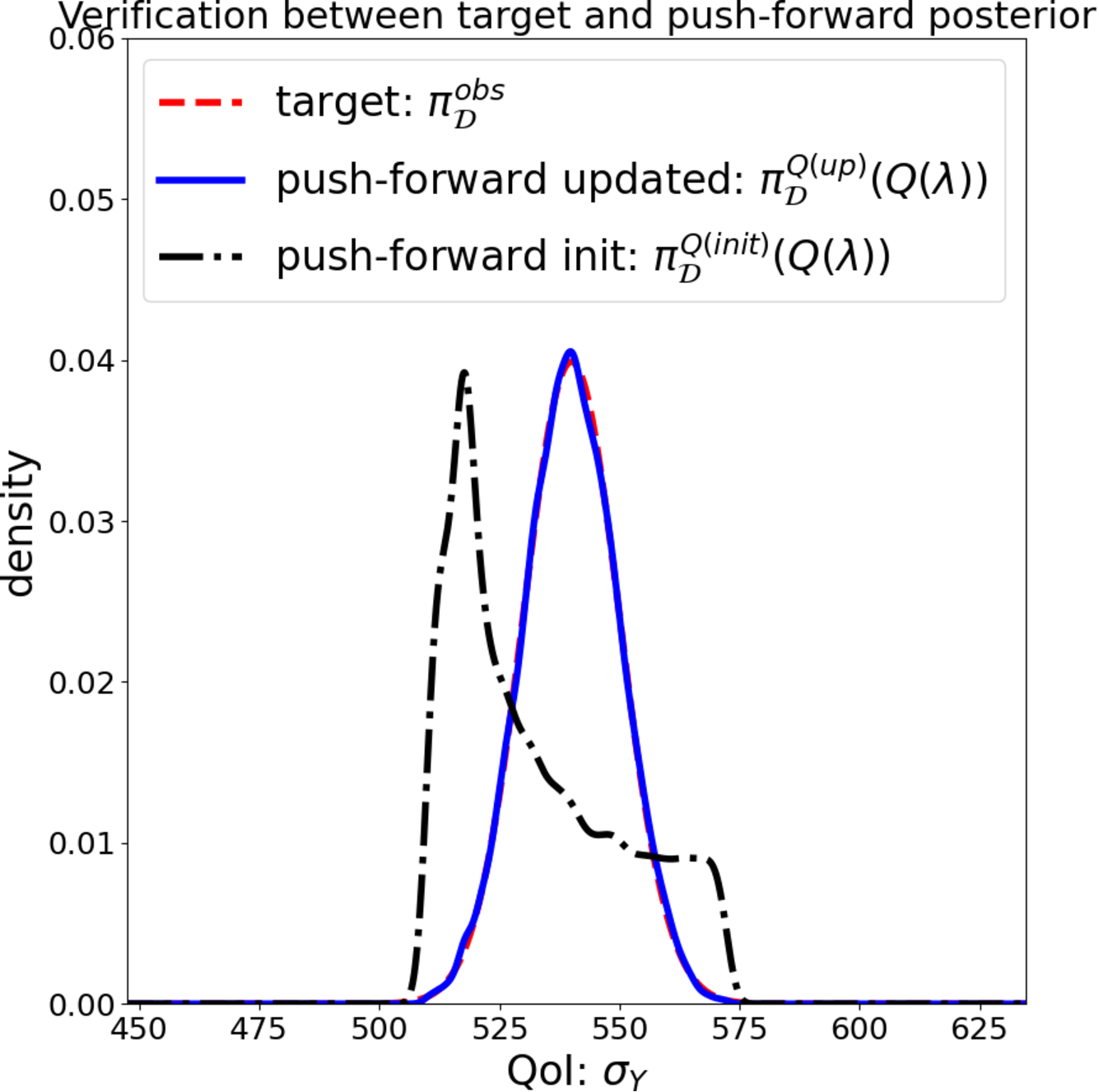}
\end{subfigure}
\caption{On the left, the initial density, $\priordens(\param) = \mathcal{U}(0.25,2.75)$, and updated density, $\postdens(\param)$, on the microstructure feature $\param = \mu_D$.
On the right, the target density on material properties, $\obsdens = \mathcal{N} (540, 10)$, and push-forwards of the initial and updated densities.
}
\label{fig:posteriorNormal-540-10}
\end{figure}

\begin{figure}[!htbp]
\centering
\begin{subfigure}[b]{0.475\textwidth}
\includegraphics[height=190px,keepaspectratio]{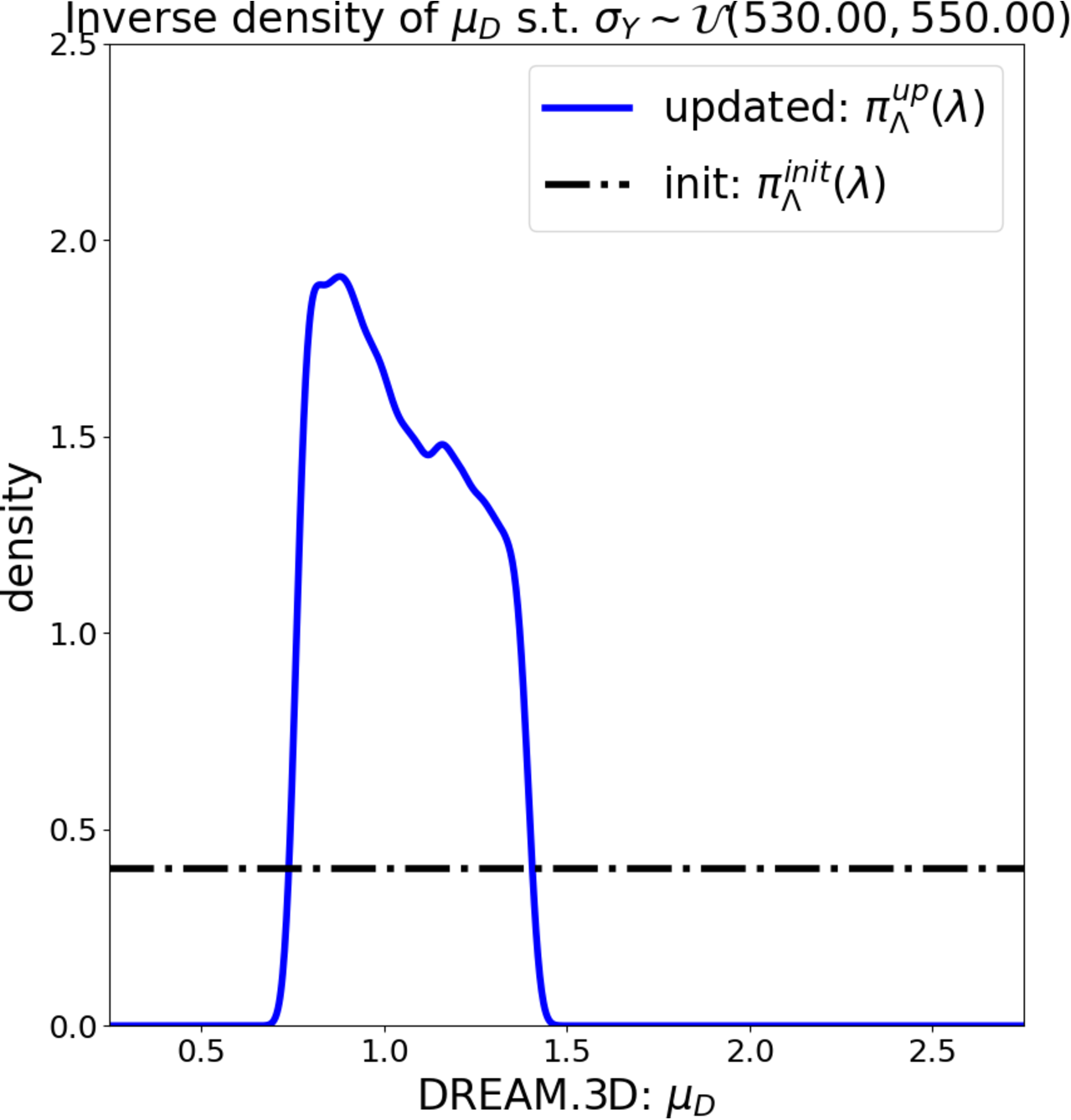}
\end{subfigure}
\hfill
\begin{subfigure}[b]{0.475\textwidth}
\includegraphics[height=190px,keepaspectratio]{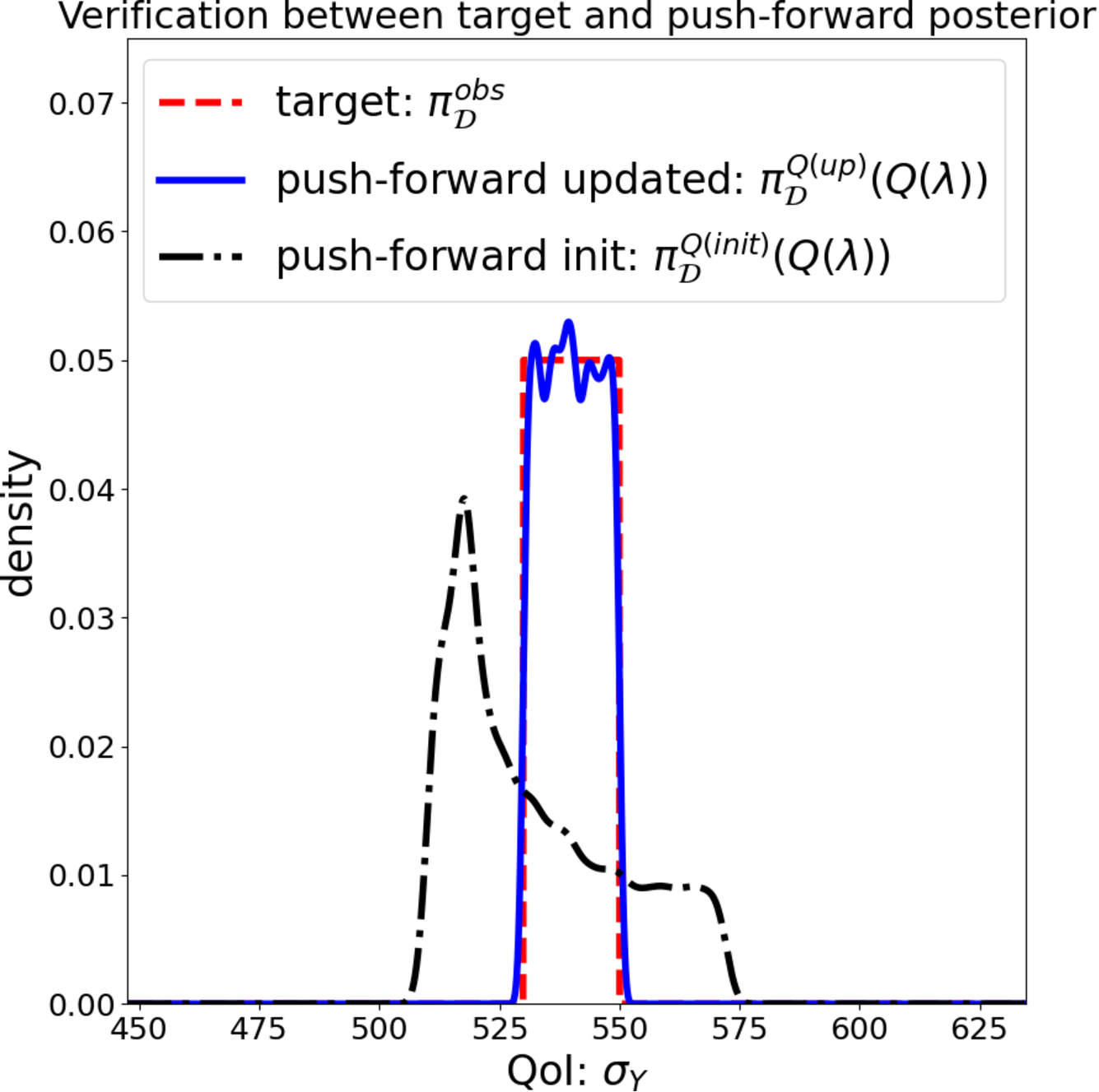}
\end{subfigure}
\caption{On the left, the initial density, $\priordens(\param) = \mathcal{U}(0.25,2.75)$, and updated density, $\postdens(\param)$, on the microstructure feature $\param = \mu_D$.
On the right, the target density on material properties, $\obsdens = \mathcal{U}(530,550)$, and push-forwards of the initial and updated densities.}
\label{fig:posteriorUnif-530-550}
\end{figure}

Using the GP regression model, two scenarios are considered to demonstrate the applicability of our proposed UQ/ML stochastic inversion framework.
For demonstration purposes, we consider two different target densities.
First, we consider the normal distribution $\sigma_Y \sim \mathcal{N} (540, 10)$, as shown in Figure~\ref{fig:posteriorNormal-540-10} (right) 
and then a uniform distribution $\sigma_Y \sim \mathcal{U} (530, 550) $, as shown in Figure~\ref{fig:posteriorUnif-530-550} (right). 
We assume a uniform initial density for $\param$, i.e., $\param \sim \priordens(\param) = \mathcal{U}(0.25,2.75)$, and use Algorithm~\ref{alg:rejectionSamplingStochInvPS} to generate $10^5$ samples from the updated density, $\postdens(\param)$.
For visualization purposes, we use a kernel density estimation (KDE) technique to construct an approximation of the updated density, but we emphasize that this KDE is not used in the rejection sampling algorithm or in the computation of the diagnostics.
Figure~\ref{fig:posteriorNormal-540-10} (left) shows the updated density of $\param = \mu_D$ characterizing the density of microstructure feature $\mu_D$ related to the average grain size $D$.
In Figure~\ref{fig:posteriorNormal-540-10} (right), we compare the target density of yield stress $\obsdens$ with the push-forward of the initial density and the push-forward of the updated density.
We see that the push-forward of the updated density matches the target quite well.
The mean and standard deviation of the push-forward of the updated density are approximated to be 540.01 and 9.90 respectively, which agree reasonably well with the corresponding values for the observed density.
The integral of the updated density, computed using~\eqref{eq:meanup}, and~\eqref{eq:kldiv}, is 0.995, which indicates that the model can predict the data and that the updated density is a probability density.
In addition, the KL-divergence from the initial to the updated, computed using~\eqref{eq:kldiv}, is 0.629, which provides a measure of the information gained in solving the stochastic inversion problem.

In Figure~\ref{fig:posteriorUnif-530-550} (left), we plot the updated density of $\param = \mu_D$ characterizing the density of microstructure feature $\mu_D$ related to the average grain size $D$ for the case of a uniform target density.
In Figure~\ref{fig:posteriorUnif-530-550} (right), we compare the target density with the push-forward of the initial density and the push-forward of the updated density.
We see reasonably good agreement between the target density and the push-forward of the updated density.
The oscillations in the push-forward of the updated density are due to the utilization of a Gaussian kernel density estimator to approximate an approximately uniform density.
We note that since the ensemble-averaged mapping is monotonic, i.e. the predictive yield strength always increases with decreasing average grain size, and both the input and output spaces are one-dimensional, the map is a bijection and therefore the stochastic inverse problem in this case study has a unique solution.

\begin{figure}[!htbp]
\begin{subfigure}[b]{0.475\textwidth}
\centering
\includegraphics[height=190px,keepaspectratio]{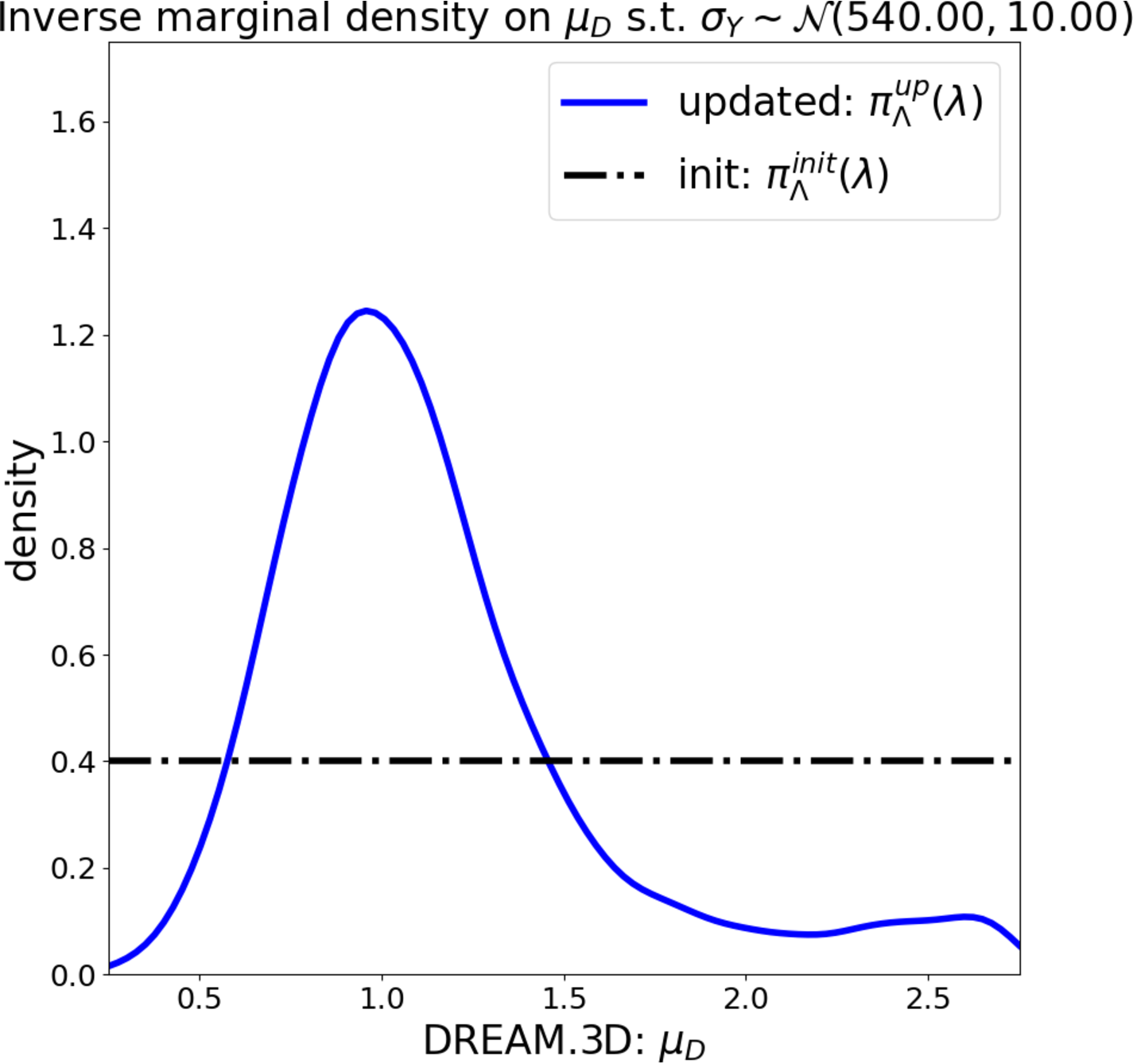}
\end{subfigure}
\hfill
\begin{subfigure}[b]{0.475\textwidth}
\centering
\includegraphics[height=190px,keepaspectratio]{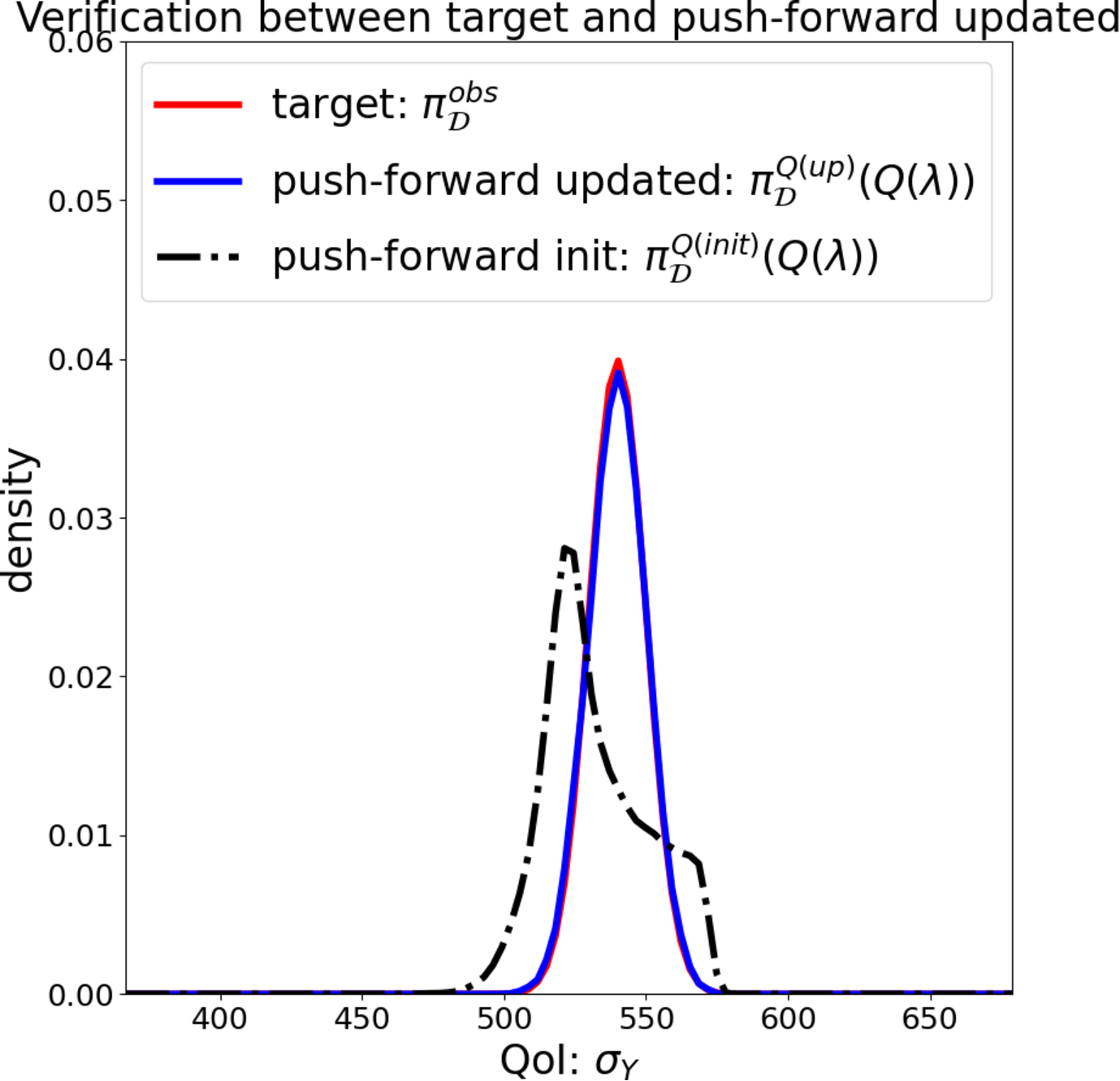}
\end{subfigure}
\caption{On the left, the initial density, $\priordens(\param) = \mathcal{U}(0.25,2.75)$, and updated density, $\postdens(\param)$, on the microstructure feature $\param = \mu_D$ using a stochastic map instead of the ensemble-averaged deterministic map.
On the right, the target density on material properties, $\obsdens = \mathcal{N} (540, 10)$, and push-forwards of the initial and updated densities.
}
\label{fig:posteriorNormal-540-10-SMAP}
\end{figure}

\black{
Next, we solve a stochastic inverse problem using the stochastic map and the heteroscedastic GP regression model in Figure~\ref{fig:smap_training_predictions}.  We focus on the case where the target density is $\sigma_Y \sim \mathcal{N} (540, 10)$.
As before, we use Algorithm~\ref{alg:rejectionSamplingStochInvPS} to generate $10^5$ samples from the updated density and plot the results in Figure~\ref{fig:posteriorNormal-540-10-SMAP}.
We clearly see that the stochasticity inherent in the map has smoothed out both the push-forward of the initial density and the updated density.  We emphasize that this updated density is really just the marginal of a consistent probability density on the joint space, which solves the stochastic inverse problem. }
The mean and standard deviation of the push-forward of this consistent density are 540.1 and 10.03 which match the the corresponding values for the target density.  In addition, the integral of the updated density is 0.994 and the KL-divergence from the initial to the updated is 0.557.  The fact that the KL-divergence is smaller for the stochastic map is consistent with the observation that the updated density is smoothed out compared with the updated density from the deterministic map.

\section{Case study 2: Aluminum with varying aspect ratios under uniaxial tension}
\label{sec:casestudy-Al}


Inspired by the studies of Liu et al~\cite{liu2020microstructure,liu2020strategy}, this case study seeks to assess the microstructure effects, specifically the grain aspect ratio, on the yield stress of an aluminum alloy.

\subsection{Constitutive models and materials parameters}


A phenomenological constitutive model used for fcc aluminum was first proposed by Hutchinson~\cite{hutchinson1976bounds} and extended for twinning by Kalidindi~\cite{kalidindi1998incorporation}.
The plastic component is parameterized in terms of resistance $\xi$ on $N_{\text{s}}$ slip and $N_{\text{tw}}$ twin systems as discussed in detail in Section 6.2.2 in Roters et al.~\cite{roters2019damask}.
We briefly summarize this model here for the sake of completeness.
The resistance on $\alpha=1,\dots,N_{\text{s}}$ slip systems evolve from $\xi_0$ to a system-dependent saturation value and depend on shear on slip and twin systems according to
\begin{equation}
\dot{\xi}^{\alpha} = h_0^{\text{s-s}} (1 + h^{\alpha}_{\text{int}}) \left[ \sum_{\alpha' = 1}^{N_{\text{s}}} | \dot{\gamma}^{\alpha'} | \left| 1 - \frac{\xi^{\alpha'}}{\xi^{\alpha'}_{\infty}} \right|^a \text{sgn}\left( 1 - \frac{\xi^{\alpha'}}{\xi^{\alpha'}_{\infty}} \right) h^{\alpha \alpha'} \right],
\end{equation}
where 
$h$ denotes the components of the slip-slip and slip-twin interaction matrices,
$h_0^{\text{s-s}}$, $h_{\text{int}}$, $c_1$, $c_2$ are model-specific fitting parameters and $\xi_{\infty}$ represents the saturated resistance evolution.

Shear on each slip system evolves at a rate of
\begin{equation}
\dot{\gamma}^{\alpha} = (1 - f^{\text{tot}}_{\text{tw}}) \dot{\gamma_0}^{\alpha} \left| \frac{\tau^{\alpha}}{\xi^{\alpha}} \right|^n \text{sgn}(\tau^{\alpha}).
\end{equation}
Following previous studies~\cite{roters2019damask,wicke2019mixed,han2020using,zhao2008investigation}, we utilize the values of the model parameters listed in Table~\ref{tab:AlumParams}.

\begin{table}[!htbp]
\centering
\caption{Aluminum constitutive model parameters (cf. Table 2 in Roters et al \cite{roters2019damask}).}
\label{tab:AlumParams}
\begin{tabular}{|l|l|l|l|l|l|} \hline
\textbf{Property}   & \textbf{Value}  & \textbf{Unit}     & \textbf{Property}         & \textbf{Value}  & \textbf{Unit}      \\ \hline
                    &                 &                   & $\tau_0$                  & 31.0            & MPa  \\
$C_{11}$            & 106.75          & GPa               & $\tau_{\infty}$           & 63.0            & MPa  \\
$C_{22}$            & 60.41           & GPa               & $a$                       & 2.25            &      \\
$C_{44}$            & 28.34           & GPa               & $h_0$                     & 75.0            & MPa  \\
$\dot{\gamma}_0$    & 0.001           & s$^{-1}$          & $h^{\alpha \beta'}$       & 1.0 or 1.4      &      \\ \hline

\end{tabular}
\end{table}

\subsection{Forward UQ for materials behaviors}

\begin{figure}[!htbp]
\begin{subfigure}[b]{0.425\textwidth}
\centering
\includegraphics[height=190px,keepaspectratio]{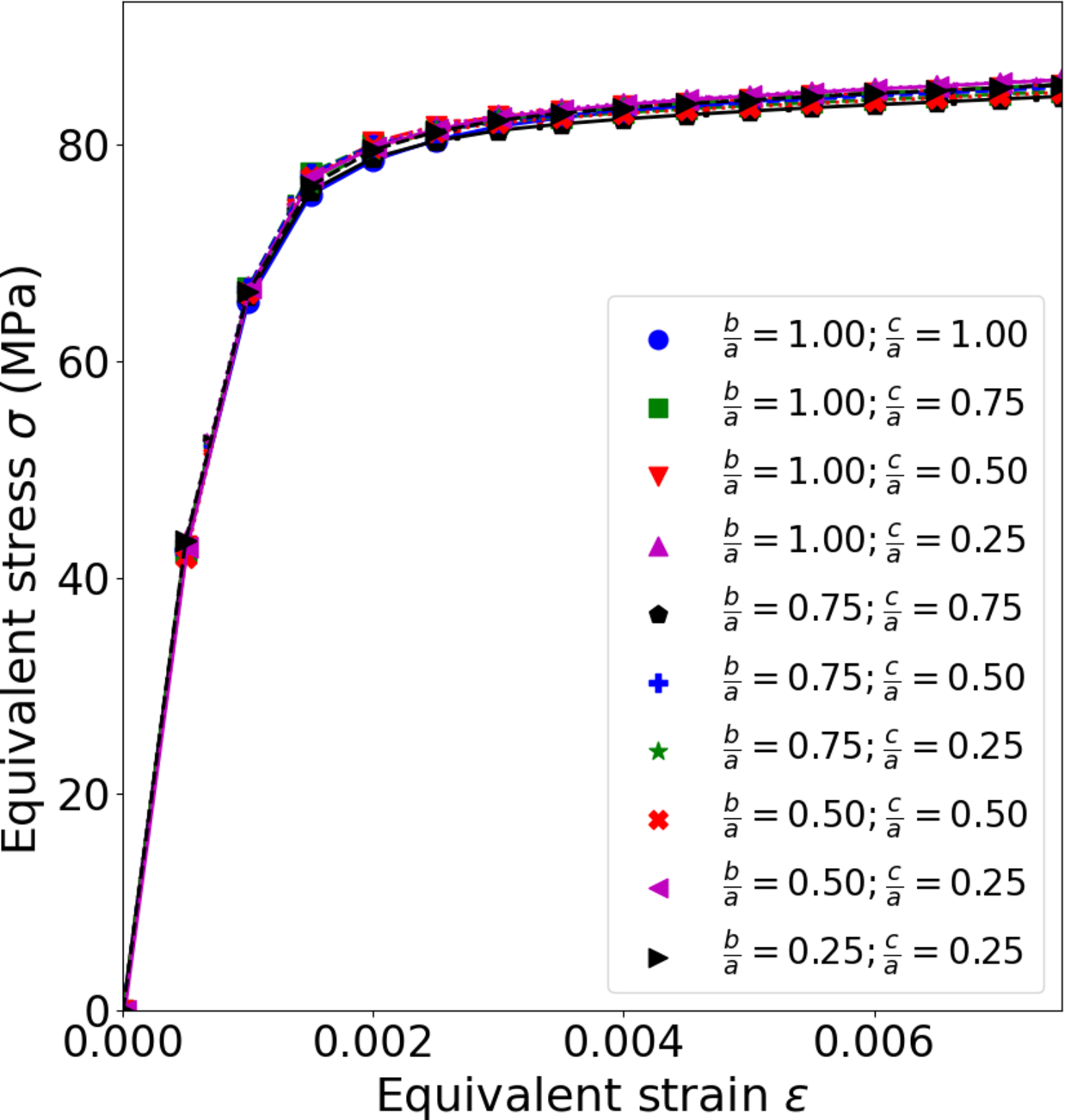}
\end{subfigure}
\begin{subfigure}[b]{0.425\textwidth}
\centering
\includegraphics[height=190px,keepaspectratio]{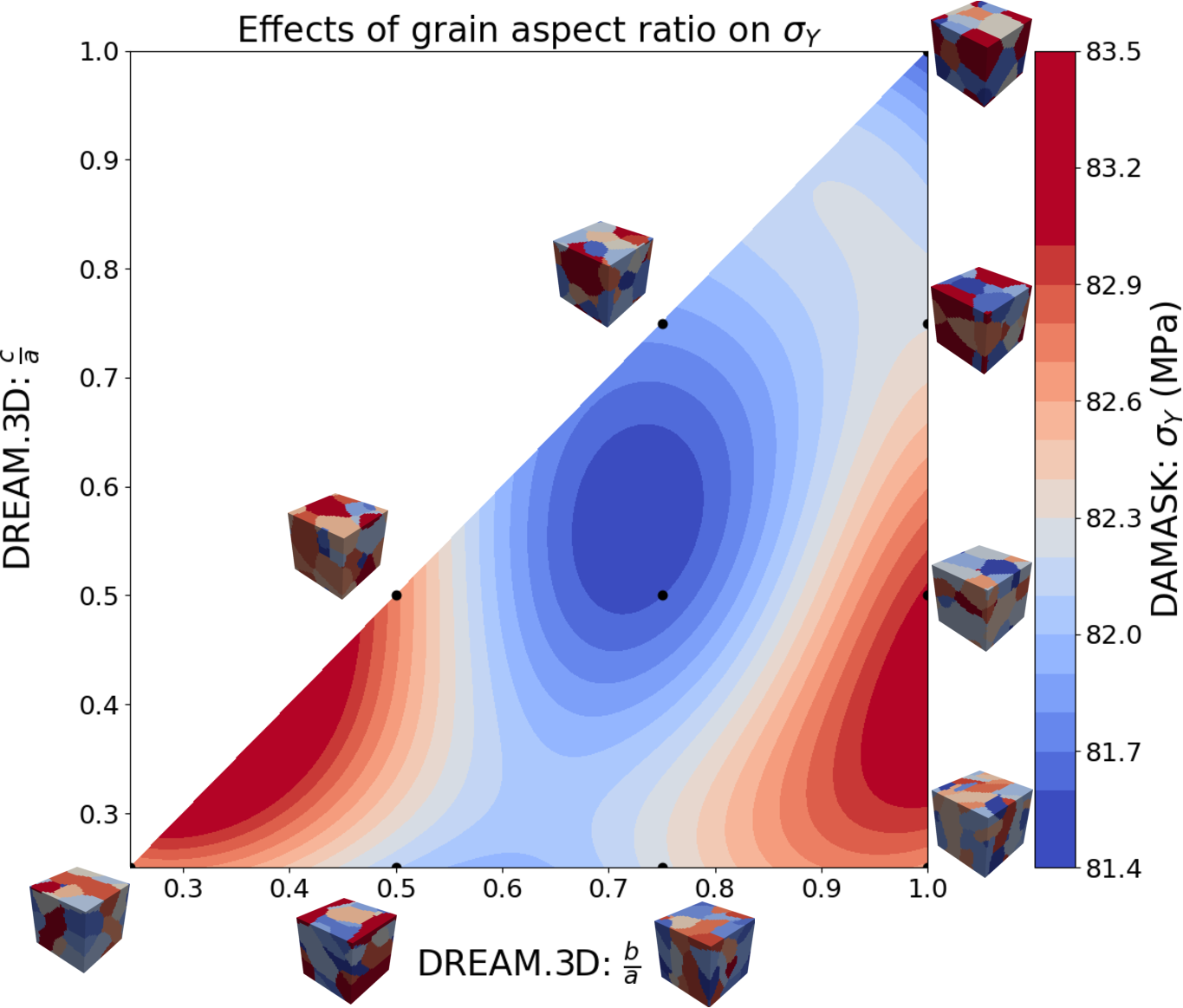}
\end{subfigure}
\caption{On the left, the equivalent stress-strain curve predicted using CPFEM model for an aluminum alloy with $36\mu m \times 36\mu m \times 36\mu m$ at $\dot{\varepsilon}_{11} = 0.001 \text{s}^{-1}$.
On the right, a contour plot showing the effect of grain aspect ratio $\left( \frac{b}{a}, \frac{c}{a} \right)$ on $\hat{\sigma}_Y$.
}
\label{fig:cropped_equivStressStrain_grain_AR_Alum}
\end{figure}

Let $a$, $b$, $c$ denote the ordered dimensions of fitted ellipsoids to a grain, i.e., $a > b > c$.
In this case study, we fix the largest dimension,$a$, and vary the aspect ratios $\frac{b}{a}$ and $\frac{c}{a}$,
i.e., $\param = \left( \frac{b}{a}, \frac{c}{a} \right)$.
We do not consider twinning in this demonstration.
For each set of aspect ratios $\left( \frac{b}{a}, \frac{c}{a} \right)$, we generate 50 SVEs using DREAM.3D, with each representing a physical region with volume of $36\mu m^3$ and discretized on a $36 \times 36 \times 36$ grid with periodic boundary conditions.
The average grain size is $9.97\mu$m and the grain size parameters are $\mu_D = 2.30$ and $\sigma_D = 0.40$.
following by CPFEM using DAMASK.
The QoI is the offset yield stress, $\hat{\sigma}_Y$, whcih is computed using DAMASK at $\varepsilon = 0.002$ under a uniaxial tension loading of $\varepsilon_{11} = 0.001 \text{s}^{-1}$.
A GP regression model is then employed as a ML tool to bridge the structure-property relationship by approximating $\sigma_Y: [0,1]^2 \mapsto \mathbb{R}$ where the ensemble average homogenized yield stress is a function of grain aspect ratio $\sigma_Y = \sigma_Y \left( \frac{b}{a}, \frac{c}{a} \right)$.
To assess the grain aspect ratio effect, we sample $\lambda = \left( \frac{b}{a}, \frac{c}{a} \right)$ at 
(1.00, 1.00), 
(1.00, 0.75), 
(1.00, 0.50), 
(1.00, 0.25), 
(0.75, 0.75), 
(0.75, 0.50), 
(0.75, 0.25), 
(0.50, 0.50), 
(0.50, 0.25), 
(0.25, 0.25), respectively. 
Figure~\ref{fig:cropped_equivStressStrain_grain_AR_Alum} (left) shows the mild effects of aspect ratio on the stress-strain curve, while Figure~\ref{fig:cropped_equivStressStrain_grain_AR_Alum} the points in the parameter space where the data is generated to train the GP model and a contour plot using the GP model.

\subsection{Inverse UQ for microstructure features}

\begin{figure}[!htbp]
\begin{subfigure}[b]{0.475\textwidth}
\includegraphics[height=190px,keepaspectratio]{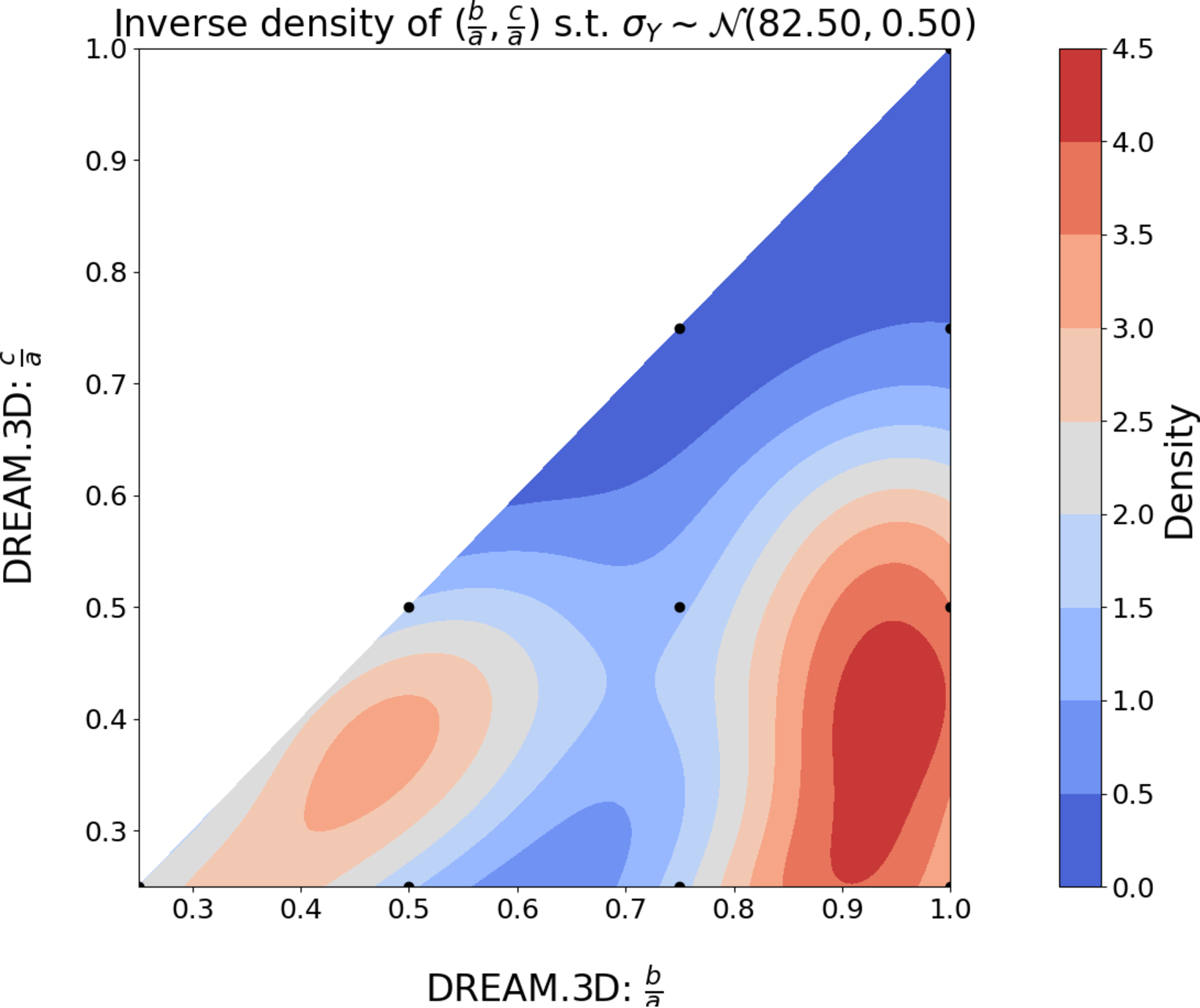}
\end{subfigure}
\hfill
\begin{subfigure}[b]{0.475\textwidth}
\centering
\includegraphics[height=190px,keepaspectratio]{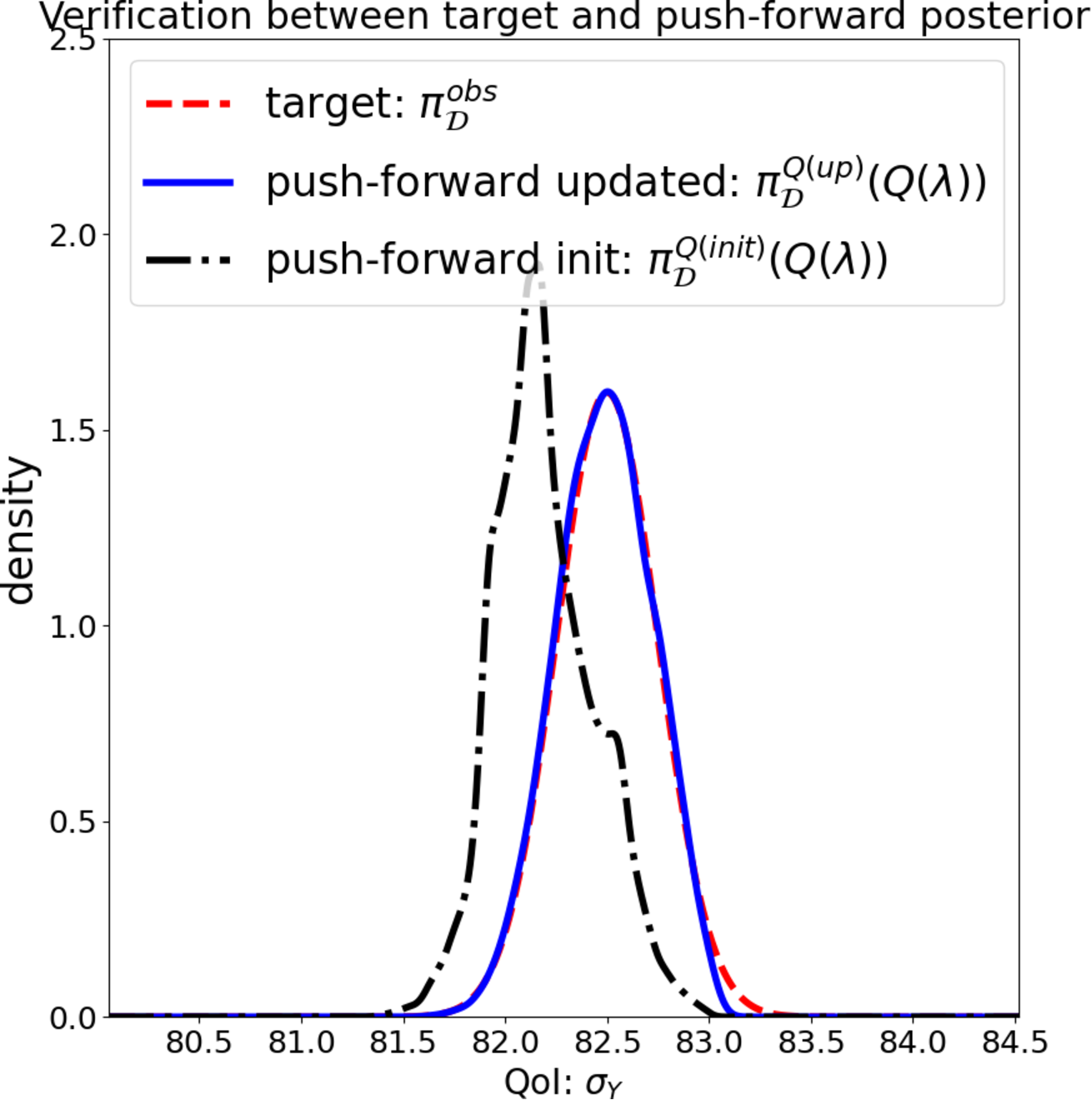}
\end{subfigure}
\caption{On the left, a contour of the updated density, $\postdens(\param)$, on the microstructure features $\param = \left( \frac{b}{a},\frac{c}{a} \right)$.
On the right, the target density on material properties, $\obsdens = \mathcal{N}(82.5,0.50)$, and push-forwards of the initial and updated densities.
}
\label{fig:posteriorNormal-82.50-0.25}
\end{figure}

\begin{figure}[!htbp]
\begin{subfigure}[b]{0.475\textwidth}
\centering
\includegraphics[height=190px,keepaspectratio]{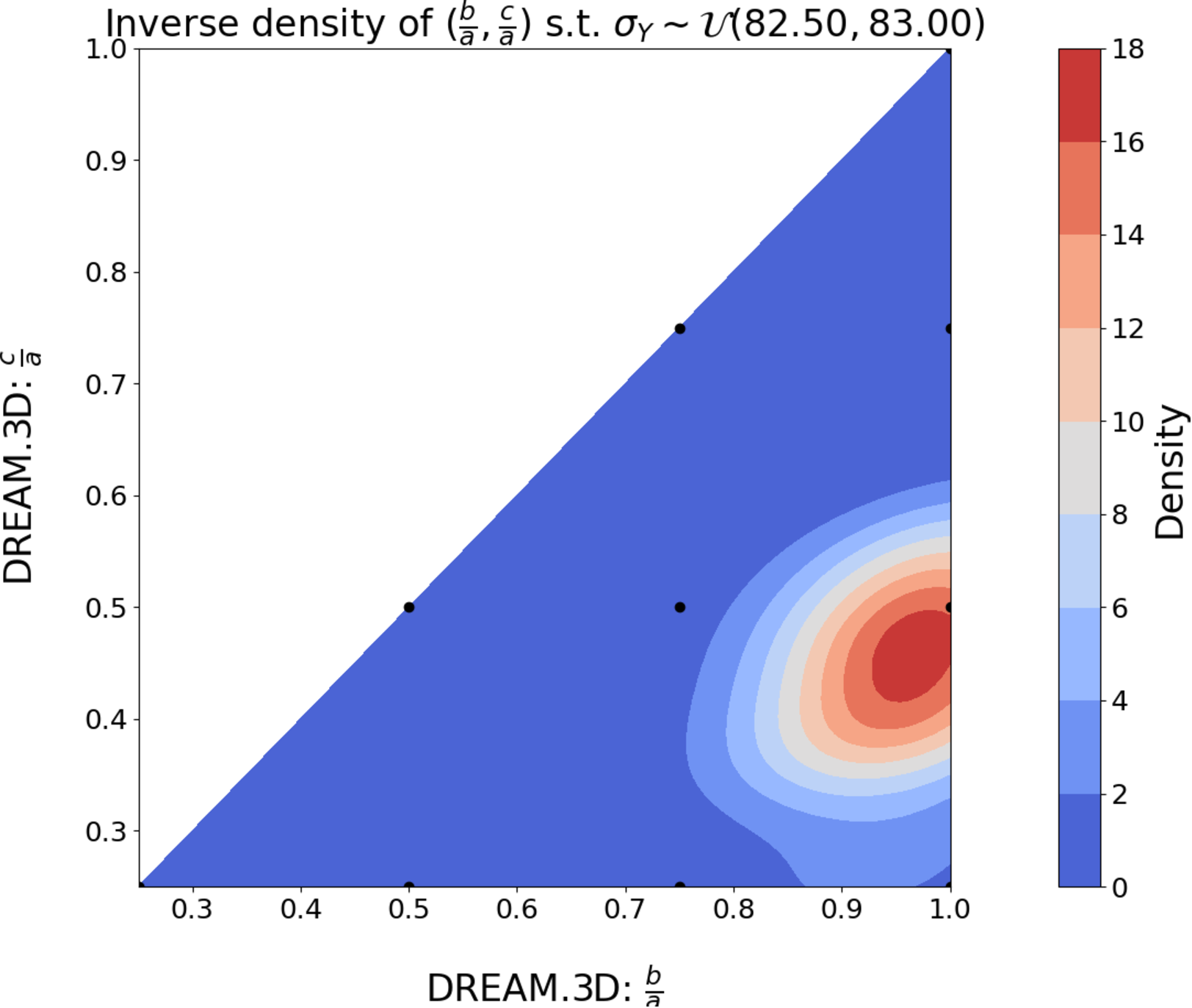}
\end{subfigure}
\hfill
\begin{subfigure}[b]{0.475\textwidth}
\centering
\includegraphics[height=190px,keepaspectratio]{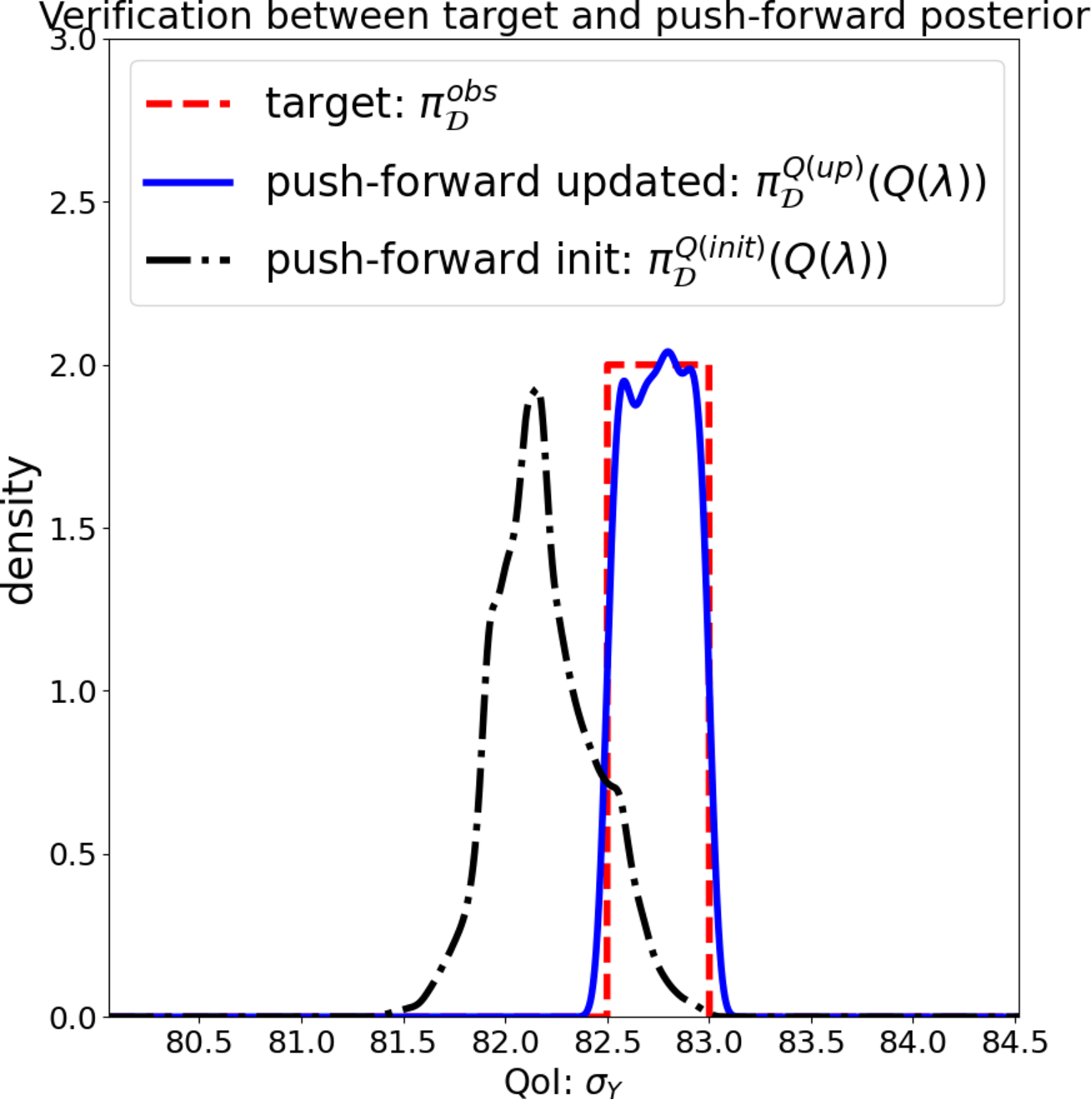}
\end{subfigure}
\caption{On the left, a contour of the updated density, $\postdens(\param)$, on the microstructure features $\param = \left( \frac{b}{a},\frac{c}{a} \right)$.
On the right, the target density on material properties, $\obsdens = \mathcal{U}(82.5,83.0)$, and push-forwards of the initial and updated densities.
}
\label{fig:posteriorUnif-8250-8300}
\end{figure}

Using the GP surrogate model constructed from the dataset described in the previous section, two scenarios are considered to demonstrate our proposed UQ/ML framework.
The target densities considered are a normal distribution $\sigma_Y \sim \mathcal{N} (82.5, 0.5)$, as shown in 
Figure~\ref{fig:posteriorNormal-82.50-0.25} (right) 
and a uniform distribution $\sigma_Y \sim \mathcal{U} (82.5, 83.0) $, as shown in 
Figure~\ref{fig:posteriorUnif-8250-8300} (right)
.
The input parameters have a natural ordering, i.e., $0.25 \leq \frac{b}{a} \leq \frac{c}{a}$, so the input domain is the triangular region depicted in 
Figure~\ref{fig:cropped_equivStressStrain_grain_AR_Alum} (left) 
and we assume a uniform initial density on this domain.
Then, we use Algorithm~\ref{alg:rejectionSamplingStochInvPS} to generate $2 \cdot 10^5$ samples from the updated density
$\postdens(\param)$.
Again, we use a standard KDE to construct an approximation of $\postdens(\param)$ strictly for visualization purposes.
\black{
Figure~\ref{fig:posteriorNormal-82.50-0.25} (left) shows the updated density over $\pspace$ and Figure~\ref{fig:posteriorNormal-82.50-0.25} (right) compares the target density of yield stress and the push-forward of the updated density $\pfpostdens(\qmap(\param))$.
We see that the push-forward of the updated density matches the target density quite well and this is confirmed numerically by computing the mean and standard deviation of the push-forward of the updated density which are estimated to be 82.49 and 0.473 respectively. 
These values agree reasonably well with the corresponding values for the observed density.
In addition, the integral of the updated density and the KL-divergence from the initial to the updated, computed using~\eqref{eq:meanup} and~\eqref{eq:kldiv}, are 0.977 and 0.635 respectively.
As previously mentioned, these numerical diagnostics are computed using only the information gathered from applying Algorithm~\ref{alg:rejectionSamplingStochInvPS}, and indicate that the updated density is actually a probability density and that some information has been learned by solving this stochastic inverse problem.
In Figure~\ref{fig:posteriorNormal-82.50-0.25}, we see that the stochastic inverse framework identifies two modes over the parameter space that are consistent with the target density.
}

Next, we consider the case of a uniform target density over a subinterval in $\qspace$.
\black{
Figure~\ref{fig:posteriorUnif-8250-8300} (left) shows the updated density on $\pspace$.
Figure~\ref{fig:posteriorUnif-8250-8300} (left) compares the target density of yield stress $\obsdens$ and the push-forward of the initial density and the push-forward of the updated density $\pfpostdens(\qmap(\param))$.
We see a relatively good match between the two densities and the oscillations are simply due to the use of the Gaussian KDE to visualize the push-forward density.
In terms of the diagnostics, the integral of the updated density is 0.994, which indicates that it is a probability density, and the KL-divergence from the initial to the updated density is 0.641, which provides an indication that information has been learned through the stochastic inversion.
Comparing Figure~\ref{fig:posteriorUnif-8250-8300} (left) and Figure~\ref{fig:posteriorNormal-82.50-0.25} (left), we see a collapse of one of the modes in the updated density.
}

\section{Discussion}
\label{sec:Discussion}

\black{
In this work, we solve stochastic inverse problems in structure-property linkages using both deterministic and stochastic maps.
By solving a stochastic inverse problem, rather than a deterministic one, the notion of the optimal microstructure is now relaxed and the updated distribution of microstructures provides a probabilistic characterization over a broader population of microstructures.
Utilizing the stochastic map, as in our first case study of TWIP steels, allows one to incorporate uncertainty from various sources.
The main difference is fully explained in recent work~\cite{butler2020stochastic}, where the solution of the stochastic inverse problem over the parameter space is interpreted as the marginal of a data-consistent solution of the stochastic inverse problem with the stochastic map.
}

In general, the stochastic inversion approach used in this paper can be applied whenever the target distribution is contained within the support of the push-forward of the initial density.
For the sake of simplicity, we focused on uniform and normal distributions for the target materials properties because these two distributions are among the most commonly used distributions for practitioners and engineers.
Exploring different target distributions is fairly straightforward and does not introduce much computational effort since the main computational cost is computing the push-forward of the initial density.

One of the drawbacks of the stochastic inversion framework is the reliance on probability densities.
We only require a mechanism to generate samples from the initial density, so the initial and updated densities do not need to be approximated.
On the other hand, the target and push-forward of the initial must be approximated in order to generate samples from the updated density.
In this paper, we employ basic Monte Carlo sampling with a GP surrogate model because this allows us to generate a large number of samples from the push-forward of the initial density for a relatively cheap computational cost, which enables the use of non-parametric density estimation for the push-forward of the initial density.
More advanced UQ approaches, such as stochastic collocation \cite{tran2019quantifying}, polynomial chaos expansions, adaptive sampling techniques, or multi-fidelity sampling \cite{tran2019sbfbo2cogp,tran2020smfbo2cogp}, could also be considered for approximating this push-forward density. 

\black{For the process-structure relationship (which is beyond the scope of this paper), where the optimal process parameters can be rightly determined, one can formulate the inverse problem in the process-structure relationship as an optimization problem and use microstructure descriptors to measure differences between microstructures as objectives. Interested readers are referred to one of our recent work \cite{tran2020an}, where Bayesian optimization is used to determine the process parameters given the microstructures.}

Generally speaking, model-form uncertainty is ubiquitous in computational science.  This is particularly true with CPFEM models where the high-fidelity dislocation-based constitutive models are much more expensive to simulate that the phenomenological constitutive models. If model-form uncertainty is quantified, i.e. the model is not perfect but one can quantify its reliability, then, depending on the characterization of the uncertainty as either epistemic or aleatoric, we can still formulate and solve a stochastic inverse problem using either the deterministic or stochastic approach respectively.

\section{Conclusion}
\label{sec:Conclusion}

\black{
This paper has presented an approach for solving stochastic inverse problems in the context of structure-property linkages that incorporates surrogate-based machine learning and measure-theoretic stochastic inversion.
The forward ML model was used to ease the computational burden in bridging the structure-property relationship.
Given this ML model, the stochastic inverse sought to infer a probability density on microstructure features that is consistent with the model and the data in the sense that the forward propagation of this probability density through the model matches a given target density on material properties.
The methodology was demonstrated using two case studies of a TWIP steel and an aluminum alloy.
Other approaches for solving inverse problems in structure-property linkages were based on optimization and gave deterministic optimal microstructure parameters.
In contrast, the methodology demonstrated in this paper relaxes this notion of optimality and the resulting updated distribution of microstructures gives a probabilistic characterization over population of microstructures.
For the sake of clarity, we have only utilized classical Monte Carlo sampling for the ensemble averages and simple rejection sampling (based on Monte Carlo) to generate samples from the updated density.  More advanced sampling strategies, such as adaptive importance sample, Markov chain Monte Carlo and multilevel or multi-fidelity Monte Carlo could also be employed, but we leave this subject for future work.
}

\begin{acknowledgements}

The views expressed in the article do not necessarily represent the views of the U.S. Department of Energy or the United States Government. Sandia National Laboratories is a multimission laboratory managed and operated by National Technology and Engineering Solutions of Sandia, LLC., a wholly owned subsidiary of Honeywell International, Inc., for the U.S. Department of Energy's National Nuclear Security Administration under contract DE-NA-0003525.  This research was supported by the U.S. Department of Energy, Office of Science, Early Career Research Program.

On behalf of all authors, the corresponding author states that there is no conflict of interest.
\end{acknowledgements}

%
%

\bibliographystyle{spmpsci}      
\bibliography{lib}   

\end{document}